\documentstyle[proceedings,epsf]{crckapb}

\newcommand{\lta}{\raisebox{-0.5ex}{$\stackrel{<}{\sim}$}}
\newcommand{\gta}{\raisebox{-0.5ex}{$\stackrel{>}{\sim}$}}
\newcommand{\tcite}[1]{\protect\citeauthor{#1}~(\protect\citeyear{#1})}
\newcommand{\refcite}[1]{\protect\citeauthor{#1}~\protect\citeyear{#1}}
\newcommand{\bibargs}[2]{\protect\citeauthoryear{#1}{#2}}
\newcommand{\earth}{{\rm M}_\oplus}
\newcommand{\sun}{{\rm M}_\odot}
\newcommand{\mmsn}{\mbox{\rm MMSN}}
\newcommand{\tstop}{t_{\rm s}}
\newcommand{\pstick}{p_{\rm s}}
\newcommand{\mplanet}{m_{\rm p}}
\epsfverbosetrue

\begin{opening}
\title{The Formation of Planets}
\author{Steven P. Ruden}
\institute{University of California, Irvine\\
Department of Physics and Astronomy\\
Irvine, CA 92697\\
USA}
\end{opening}

\begin{document}
\section{Introduction}
\label{sec:intro}

Mankind has attempted to understand the origin of the Earth and planets for countless
generations.  In the past few decades,
starting largely from the pioneering work of Safronov in the 1960s,
philosophical speculation about their birth
has been replaced with detailed analytical and numerical simulations whose
goal is to reproduce the observed properties of our own solar system.
With the recent groundbreaking discoveries of planets around ordinary stars like our Sun
(\citeauthor{mayor} \citeyear{mayor}; \citeauthor{marcyb} \citeyear{marcyb}; 
\citeauthor{butlerm} \citeyear{butlerm}; see also the review by Marcy in 
this volume), we now have a reasonable sample of nearby planetary systems that we can
use to test and validate theoretical models and from which we can learn new
and unexpected features of planet formation.  
In the upcoming decade 
the launches of NASA's Space Interferometry Mission (SIM)
and ESA's Global Astrometric Interferometer for Astrophysics (GAIA)
(and their longer term counterparts Terrestrial Planet Finder [TPF]
and Infrared Space Interferometry Mission [IRSI])
will allow us to detect {\em directly} planets around nearby stars, which will give 
us an even more complete census of the number and content of planetary systems 
in the Solar neighborhood.  
To place this observational data in a proper astrophysical context, we need
to understand how planets are born and how they interact with their environment.
In this review, I will discuss the dynamics of the growth of solid particles
from initially micron-sized dust grains through kilometer-sized planetesimals 
to their final planetary mass.  I will also discuss the formation of the gas--rich
giant planets and their tidal interaction with the surrounding nebular disk.  
The reader interested in further details of the planet formation process
is encouraged to consult the excellent reviews by Lissauer~\shortcite{lisreview}
and by Marcy and Butler~\shortcite{marcybreview}.

\subsection{Overview: From Dust to Planets}
\label{sec:over}

The low eccentricities and inclinations of the planets (see Table~\ref{planettable})
have long been used as evidence that the Solar System formed from 
a flattened disk of gas and dust: the {\em primordial solar nebula}.  The 
recognition that the collapse of magnetized, rotating molecular 
cloud cores naturally leads to protostellar disks with roughly the right mass and size
($\sim 100$~AU) is one of the successes of modern star formation theory (for a thorough
review see \citeauthor{shureview}~\citeyear{shureview}).
The solid content in these disks is comprised of the interstellar grains that survive
their passage into the disk (through a surface disk shock) plus nebular volatiles
that condense as dust within the disk.  
In the current ``standard model'' of planet formation, the rocky cores of all planetary bodies
grow from the accumulation of small dust grains;
the increase in growth is tremendous -- nearly a factor of $10^{13}$ in size and $10^{40}$ in mass.  
In our Solar System, this model is sufficient to explain the formation of the 
terrestrial planets (Mercury, Venus, Earth, Mars), which have
densities comparable to rock (see Table~\ref{planettable}) and
low-mass gas atmospheres that were
most likely outgassed from the rocky material from which they accreted
(\refcite{brown}; \refcite{prinn}; \refcite{pepin}).
The giant planets (Jupiter, Saturn, Uranus, Neptune), which have much lower
mean densities ({\em cf.} Table~\ref{planettable}) and massive gas
atmospheres, are also thought to have formed from initially rocky cores\footnote{Models 
invoking dynamical
gravitational instabilities to form the giant planets (\refcite{cameron}) have fallen out of 
favor (see \citeauthor{linpapII}~\citeyear{linpapII}).}.
However, once the core mass grows to be larger than a critical value ($\gta 10 \earth$),
rapid accretion of nebular gas transforms them into gas giant planets (\refcite{mizuno}).

\begin{table}[htb]
\begin{center}
\caption{Properties of Solar System Planets}
\label{planettable}
\begin{tabular}{cccccc}
\hline
 & Orbital & & & & \\
 & Semimajor & &Mean & & \\
Planet & Axis & Mass & Density &Eccentricity & Inclination\\
 &(AU)  &($\earth$) & (g/cm$^3$)& &(degrees) \\
\hline
Mercury& 0.39 & 0.055 &5.4&0.20 & 7\\
Venus&  0.72 & 0.82 & 5.2 & 0.007 & 3\\
\hspace*{6mm}Earth ($\earth$) & 1.0 & 1.0 & 5.5 & 0.02 & 0\\
Mars & 1.5 & 0.11 &4.0 & 0.09 & 1\\
Jupiter & 5.2 & 318 & 1.3 & 0.05 & 1\\
Saturn & 9.6 & 95.1 & 0.70 & 0.06 & 2\\
Uranus & 19 & 14.5 & 1.6 & 0.05 & 1\\
Neptune &30 & 17.2 & 1.6 & 0.009 & 2\\
\hline
Notes:&\multicolumn{5}{l}{\parbox[t]{3.5in}{Pluto is not included because it is more typical
of the low--mass outer Solar System Kuiper belt objects.\\
$1\earth = 6 \times 10^{27} {\rm g} = 3 \times 10^{-6} \sun$}}
\end{tabular}
\end{center}
\end{table}

The growth of planet-sized
solid bodies conventionally occurs in two loosely defined sequential phases.
In the first phase (\S\ref{sec:grainplanetesimal}), 
dust grains, which are uniformly distributed in and strongly coupled
to the nebular gas, grow through binary collisions.  As they grow larger they settle 
gravitationally towards the nebular midplane forming a thin layer.
Mutual collisions (and possibly gravitational instability of the dust layer)
induce further growth in the midplane forming
roughly kilometer-sized bodies called {\em planetesimals}.
Planetesimals are, by definition, massive enough to have largely decoupled from the gas
and move on nearly circular, Keplerian orbits about the central star.  The first phase is believed
to be relatively short, occurring on a timescale as little as a few thousand orbital periods
(\refcite{stucuzzi}).
In the second phase, planetesimals continue to grow through inelastic collisions.
The essential feature is that collision cross-sections are increased above their
geometrical value by the focusing nature of the gravitational interaction between the
colliding planetesimals (\refcite{safronov}).
If the relative velocity dispersion among the planetesimals
remains sufficiently small,
the increase in cross-section is greatest for the most massive planetesimal.
This planetesimal will then collide more frequently and gain mass more rapidly than
its neighbors, which causes its collision cross-section to increase even further.
``Runaway'' growth occurs in which the most massive planetesimal accretes all the available solid
matter in its local ``accretion zone'' -- the region of the disk over which its gravity can perturb
nearby planetesimals into colliding orbits.
The runaway process produces a series of rocky protoplanetary bodies that are dynamically
and physically isolated from each other and occurs on timescales of $\sim 10^6$ years
(\citeauthor{wetherstew}~\citeyear{wetherstew}; \refcite{aarseth};
\citeauthor{stuplanet}~\citeyear{stuplanet}).
After the runaway stalls because all the available nearby solid matter
has been accreted, further collisional 
growth of the protoplanets can occur but only on much longer timescales ($\sim 10^8$ years)
as mutual gravitational perturbations cause orbits to cross and highly inelastic
collisions to occur (\citeauthor{wether90}~\citeyear{wether90}; 
\citeauthor{chambers}~\citeyear{chambers}).

As the planets grow, they find themselves in 
a nebular environment 
that changes with time as the gas in the disk is accreted by the central star
(\refcite{rudlin}).
Eventually, the disk is cleared of gas via processes such as
stellar winds or UV irradiation by the central star (\refcite{shuevap}).
The growth of the terrestrial planets to their current masses could have 
continued long after the dispersal of the gas disk.
However, because the more massive atmospheres of the giant planets were accreted from
the nebular gas disk, the growth of their rocky cores to the critical mass 
had to be complete {\em before}
disk dispersal.  Observations indicate
protostellar disk lifetimes are $\lta 10^7$ years (\citeauthor{stromdisk}~\citeyear{stromdisk}; 
\citeauthor{beckdisk}~\citeyear{beckdisk}; \refcite{zuckerman};
see also the review by Beckwith in this volume), 
placing an important time constraint on the planet formation process.
Current models have difficulty forming the giant planet cores on this timescale unless
the mass of the solar nebula is significantly larger than the minimum-mass solar nebula.

The giant planets are massive enough that gravitational tidal interactions
between the planet and the surrounding disk can create annular {gaps} largely devoid
of gas at the orbit of the planet.  This mechanism of {\em tidal truncation} 
(\refcite{linpapIII}) has been invoked to explain why the giant planets stop their
rapid gas accretion and thus attain their final masses (\S\ref{sec:tides}).  The 
tidal interaction also causes the orbital semimajor axes of the planets to
decrease in time.  This {\em orbital migration} of the giant planets is a serious
problem that will be discussed in more detail in \S\ref{sec:migrate}.

\subsection{Future Research Questions}

Throughout this review I will list a selection of the major unsolved questions in
planet formation theory.  I hope these questions will
provoke and encourage the reader to consider devoting some thought to their 
solution.

\begin{itemize}
\item When does the planet building epoch begin?
        \begin{itemize}
        \item Is it a continuous process beginning from the initial formation
                of the disk?
        \item Does planet formation require some critical conditions 
        (such as a sufficiently cool or quiescent nebula) before it begins?
        \end{itemize}
\item How does planet formation modify the disk environment and its evolution?
        \begin{itemize}
        \item Are their unambiguous observable signatures such as holes or gaps?
        \end{itemize}
\item What outcomes are likely?
        \begin{itemize}
        \item A diverse planetary system like our own?
	\item Giant planets close to the central star as in 51 Pegasus?
        \item Rubble disks like $\beta$~Pictoris?
        \end{itemize}
\end{itemize}

\section{Structure and Evolution of Disks} 
\label{sec:diskevol}
Protoplanetary disks are just one example of an accretion disk -- a geometrically
thin disk of gas (and dust) revolving around a central mass point
(\citeauthor{pringle}~\citeyear{pringle}; \citeauthor{linpapII}~\citeyear{linpapII};
and the review chapter by Kenyon in this volume).   We can understand their basic
properties by examining the force balance equations assuming the gas is in orbit
around a central protostar with mass $M_\ast$.
Adopting cylindrical
coordinates ($r,\phi,z$) and assuming axisymmetry, the disk is in vertical hydrostatic 
equilibrium where the vertical pressure force balances the vertical component of the central
star's gravity
\begin{equation}
\label{hse}
\frac{1}{\rho} \, \frac{\partial P}{\partial z} = - \frac{GM_\ast}{r^2} \,\left(\frac{z}{r}\right)
\equiv - \Omega^2 z,
\end{equation}
where $P$ and $\rho$ are the gas pressure and density, and where we have defined the
Keplerian rotation rate 
\begin{equation}
\label{keplerfreq}
\Omega = \sqrt{\frac{GM_\ast}{r^3}}.
\end{equation}
We can scale 
equation~(\ref{hse}) to find an expression for the vertical height of the nebula, $H$,
\begin{equation}
\label{scaleht}
H = \frac{c}{\Omega},
\end{equation}
where the nebular sound speed $c$ is defined by the relation $P=\rho c^2$.
We say the disk is geometrically thin if $H\ll r$, which from equation~(\ref{scaleht}) is
equivalent to the disk being dynamically
cold, $c \ll \Omega r$, {\em i.e.}, the nebular sound speed is much
less than the orbital Kepler speed.  
Because disks are thin, it is convenient to use vertically averaged quantities such
as the surface density, $\Sigma$:
\begin{equation}
\label{surden}
\Sigma = \int_{-\infty}^{+\infty} \rho \, dz \approx 2 \rho H = 2 \frac{\rho c}{\Omega},
\end{equation}
where we interpret $\rho$ in the last two terms on the right side of
the equation as the value of the density at the midplane of the nebula.

Force balance in the radial direction is given by
\begin{equation}
\label{radforce}
\frac{v^2}{r} = \frac{GM_\ast}{r^2} + \frac{1}{\rho} \frac{\partial P}{\partial r},
\end{equation}
where $v$ is the orbital speed of the gas.  For thin, cold disks, the radial pressure
gradient is smaller than the stellar gravitational force by $\sim (H/r)^2 \ll 1$ (as
may be verified by scaling eq.~[\ref{radforce}] and using eq.~[\ref{scaleht}]) and is
typically neglected in most accretion disk applications.  In this case, 
equation~({\ref{radforce}) shows the gas
rotates in centrifugal balance at the Keplerian orbital speed, $v=\Omega r$.
In most disks, the pressure is
largest near the star and decreases
outward so that the radial pressure term in equation~(\ref{radforce})
is negative.  This means the pressure of the gas partially supports the disk
against gravity, and the disk gas rotates at a slightly {\em sub-Keplerian}
speed $v= \Omega r - \Delta v$ with $\Delta v>0$.  If we substitute this expression 
for $v$ into equation~(\ref{radforce}), we find
\begin{equation}
\label{deltav}
\Delta v = \frac{1}{2} \left\vert \frac{\partial \ln P}{\partial \ln r}\right\vert
\frac{c^2}{\Omega r}
\approx \frac{c^2}{\Omega r} = \left(\frac{H}{r}\right) c \ll c,
\end{equation}
where we have used equation~(\ref{scaleht}) in the last terms on the right.
The departure from Keplerian motion is indeed very small being much less than the sound speed
which is itself much less than the Keplerian rotation speed.
We shall
see in \S\ref{sec:particlegas} that this small departure from exact Keplerian rotation
plays a critical role in the motion of solids in protoplanetary disks.

In two seminal papers, \citeauthor{LBP}~\shortcite{LBP} and \citeauthor{SS}~\shortcite{SS}
demonstrated that the evolutionary behavior of accretion disks is governed by the {\em outward}
transport of angular momentum, which allows mass to flow {\em inward} to be accreted by
the central object.  The precise mechanisms that cause the transport in {\em any} 
astrophysical disk are still far from certain.  Numerous suggestions have been made
in the case of protostellar disks (see the review by \refcite{adamslin}): 
gravitational instabilities 
(\citeauthor{LPgrav}~\citeyear{LPgrav}; \citeauthor{SLING}~\citeyear{SLING}),
magnetic instabilities (\citeauthor{steplev}~\citeyear{steplev};
\citeauthor{balbus}~\citeyear{balbus}),
thermal convective instabilities (\citeauthor{linpapconv}~\citeyear{linpapconv};
\citeauthor{rudlin}~\citeyear{rudlin}),
stellar or disk winds (\citeauthor{hartmac}~\citeyear{hartmac}; \refcite{wardle})
and fluid dynamical shear instabilities (\refcite{bebe}).
In most applications, the details of the physical cause of the transport are ignored and
angular momentum is assumed to be transported by some form of 
localized fluid dynamical turbulence that creates an eddy viscosity in the gas.  
The disk evolution is governed by the magnitude of the turbulent
viscosity, $\nu$, which is the product of the 
turbulent eddy size, $\ell_{\rm t}$, and velocity, $v_{\rm t}$.  
The eddies are assumed to move subsonically and to have sizes smaller than
the vertical height of the disk, which leads to the well-known parameterization of the viscosity
as (\refcite{SS})
\begin{equation}
\label{viscosity}
\nu \sim v_{\rm t} \cdot \ell_{\rm t} \equiv \alpha \, c \, H,
\end{equation}
where $\alpha \lta 1$ is a dimensionless parameter.
Under the action of viscous stress, an initial surface density distribution
evolves diffusively by spreading outward while transferring mass inward to be accreted 
by the central protostar.  The viscous evolutionary timescale is
\begin{eqnarray}
\label{viscoustime}
t_{\rm vis} & \approx &\frac{r^2}{\nu} \approx \frac{1}{\Omega}\,\frac{1}{\alpha}\,
\left(\frac{r}{H}\right)^2,\\
&\approx & 10^7 \left(\frac{10^{-2}}{\alpha}\right) \, \left(\frac{r/H}{25}\right)^2 
\left(\frac{r}{100 {\rm AU}}\right)^{3/2}\;\mbox{years.}\nonumber
\end{eqnarray}

The solids in protoplanetary disks find themselves in a continuously changing nebular 
environment in which temperature, density, pressure, and turbulent velocity typically
decrease with time (after transients from the initial distribution of matter decay;
see Fig.~\ref{nebmodels}).  Detailed calculations 
using reasonable values for the eddy viscosity ($\alpha \approx 10^{-2}$)
yield evolutionary
times of order $10^6 - 10^7$ years
(\citeauthor{rudlin}~\citeyear{rudlin}; \citeauthor{rudpoll}~\citeyear{rudpoll}),
which are in agreement with the observational disk lifetimes cited above.

\begin{figure}[!htb]
\epsfxsize=\hsize
\epsfbox{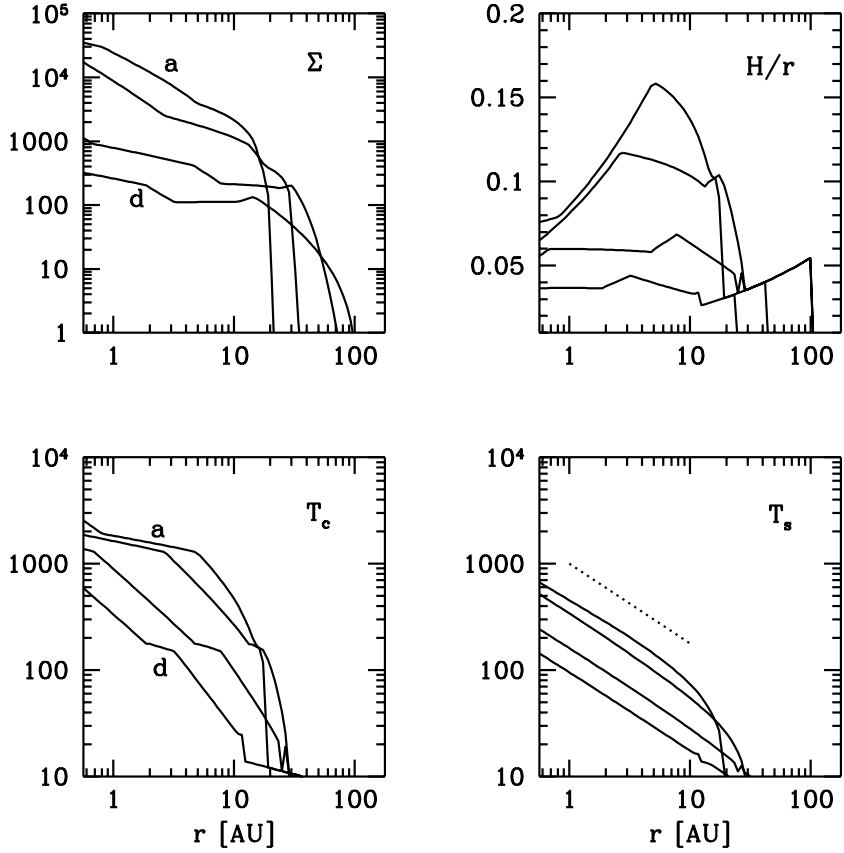}
\caption{Models for the evolution of the primordial solar nebula.  Surface density, $\Sigma$,
aspect ratio, $H/r$, central (midplane) temperature, $T_{\rm c}$, and
surface (effective) temperature, $T_s$, are plotted versus
distance from the sun (in AU) for the times
a) $10^3$, b) $10^4$, c) $10^5$, d) $10^6$ years.
Note the diffusive spreading of the nebula from less than 10~AU to 100~AU,
and the cooling of the nebula with time.  The disk always remains geometrically thin ($H/r \ll 1$),
and at late times $H/r \approx 1/25$.  The surface temperature 
falls off as $r^{-3/4}$ (marked as the dotted line) away from the outer disk edge.
The viscous $\alpha$ parameter is $10^{-2}$ in these calculations 
adapted from \tcite{rudlin}.}
\label{nebmodels}
\end{figure}

\subsection{The Minimum-Mass Solar Nebula (MMSN)}
\label{sec:mmsn}

The Solar System was formed from the collapse of a molecular cloud having 
solar element abundances, yet the
present day planets are clearly of non-solar composition.
We can ask the important cosmogonical question, ``How much {\em solar composition}
gas is required to make the planets?''  The mass and inferred spatial distribution of
this gas are known as the {\em Minimum-Mass Solar Nebula} (hereafter, \mmsn).
The answer to this question has
been given by several authors (\refcite{stu77mmsn}; \refcite{hayashi})
who take the present amount of condensed solid matter in the
Solar System and augment it to solar composition to determine the minimum mass.
The radial surface density distribution of this mass is deduced by 
smearing out the augmented mass of a planet over an
area defined by its nearest neighbors.\footnote{An important
assumption is that the planets formed at their present locations.}

A key ingredient in the analysis is the dust-to-gas ratio, $\zeta$, which
measures the mass fraction of dust ({\em i.e.}, high-$Z$ elements) 
in a solar composition gas.  In cool regions of the nebula
with temperatures below about 170~K, both rocky and icy material (such as water,
methane and ammonia) can condense from the gas  giving 
the dust-to-gas ratio a value $\zeta \approx 1/60$ (\refcite{hayashi}).
In warmer regions (above 170~K but below about 1500~K where much of the dust
evaporates) only the more refractory
(high condensation temperature) solids survive and the dust-to-gas ratio drops
by a factor of four to $\zeta \approx 1/240$ (\refcite{hayashi}).  
To find a rough value for the mass in the \mmsn\ 
we must estimate the present amount of solids in the planets.
This comes almost entirely from the high-$Z$ material inside the giant planets,
which is estimated to range from 40 - 80 $\earth$ (\refcite{hubbard};
\refcite{chabrier})
compared to only 2 $\earth$ in the terrestrial planets.  
Adopting a value of 60 $\earth$ for the present mass of condensed
solids and an average dust-gas-ratio of 1/100 gives a rough estimate of
6000 $\earth$ $\sim 0.02 \sun$ for the mass of solar composition gas needed to
make the planets.\footnote{Adding the mass of solar composition gas already present
in all the planetary atmospheres ($\lta 400 \earth$) is only a small correction.}
A more detailed calculation of the \mmsn\  by \tcite{hayashi} is illustrated in 
Figure~\ref{mmsnfig}.
Disk masses derived from millimeter observations of nearby star forming 
regions are found to range from 0.005 -- 0.2 $\sun$ (\refcite{becksarg};
also see the review by Beckwith in this volume), comparable to the \mmsn.

\begin{figure}[htb]
\epsfxsize=\hsize
\epsfbox{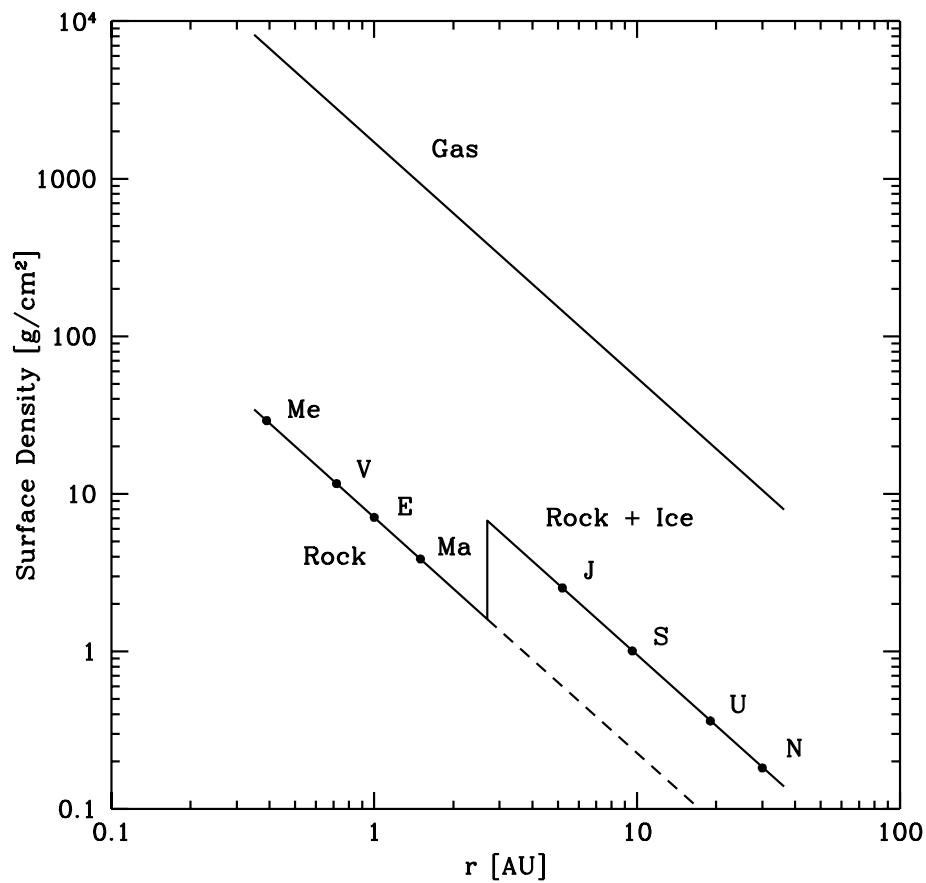}
\caption{Surface density distribution in the Minimum-Mass Solar Nebula (MMSN)
adapted from \tcite{hayashi}.  The planets are denoted by Me, V, E, Ma, J, S, U, N.
Outside 2.7~AU the temperature is deduced to be cool enough for ices to condense,
which causes the surface density of solids to jump by a factor of four.  The gas distribution
shows no discontinuity, and the surface density of
both gas and solids falls off as $r^{-3/2}$.  The total mass between
0.35~AU and 36~AU is 0.013 $\sun$.}
\label{mmsnfig}
\end{figure}

The \mmsn, while playing a useful role as a conceptual guide,
is perhaps becoming overused as a quantitative ``reference'' standard in
model calculations of the planet formation epoch.
The planets formed in an {\em evolving} disk whose properties (such 
as the position where ices condense) 
changed with time; the static $r^{-3/2}$
surface density distribution presented by the \mmsn\  may not only be
quantitatively wrong, but misleading is well.
If it does accurately describes any part of the
evolution of the solar nebula, it is probably the {\em final} stages when planet formation
is complete rather than the {\em initial} stages when planet formation began.
Furthermore, as we will see in \S\ref{sec:migrate}, 
it is now thought that orbital evolution of the giant
planet semimajor axes may have occurred, which would modify
any current reconstruction of  the \mmsn\ surface density distribution
from that depicted in Figure~\ref{mmsnfig}.

\subsection{Nonlinear Nature of Disk Physics}

It is obvious there is a very complicated interrelationship among
all the physical processes that can occur in protoplanetary disks.  Figure~\ref{nebuladeps}
is an intentionally complicated schematic diagram indicating some of the predominant
physical processes, their dependencies and their feedbacks.  Although there are many
models by many different research groups addressing the key underlying physics in
any one of the boxes in Figure~\ref{nebuladeps}, successful 
integration of the different processes together into a coherent whole has been elusive.
There are several reasons for this:
the inherent complexity of the physics involved,
the lack of experimental guidance as to the magnitude of important physical parameters,
the inability of current observational techniques to probe the relevant scales (which
occur at angular resolutions of $\sim 10^{-2}$ arcseconds in the nearest star
forming complexes) and the nonlinear interplay among the processes themselves.
For these reasons, we are far from having a theory that can predict the type of 
planetary system formed from a given nebular environment.
Furthermore, 
researchers have typically focused their modeling on explaining the properties of our
own Solar System, perhaps rejecting scenarios that, although inadequate for our 
own system,
might be good representations of the physics in other protoplanetary disks. 
As is often the case, more complete observational data will enable modelers to explain
the diversity of planetary systems found in nature.

\begin{figure}[htb]
\epsfxsize=\hsize
\epsfbox{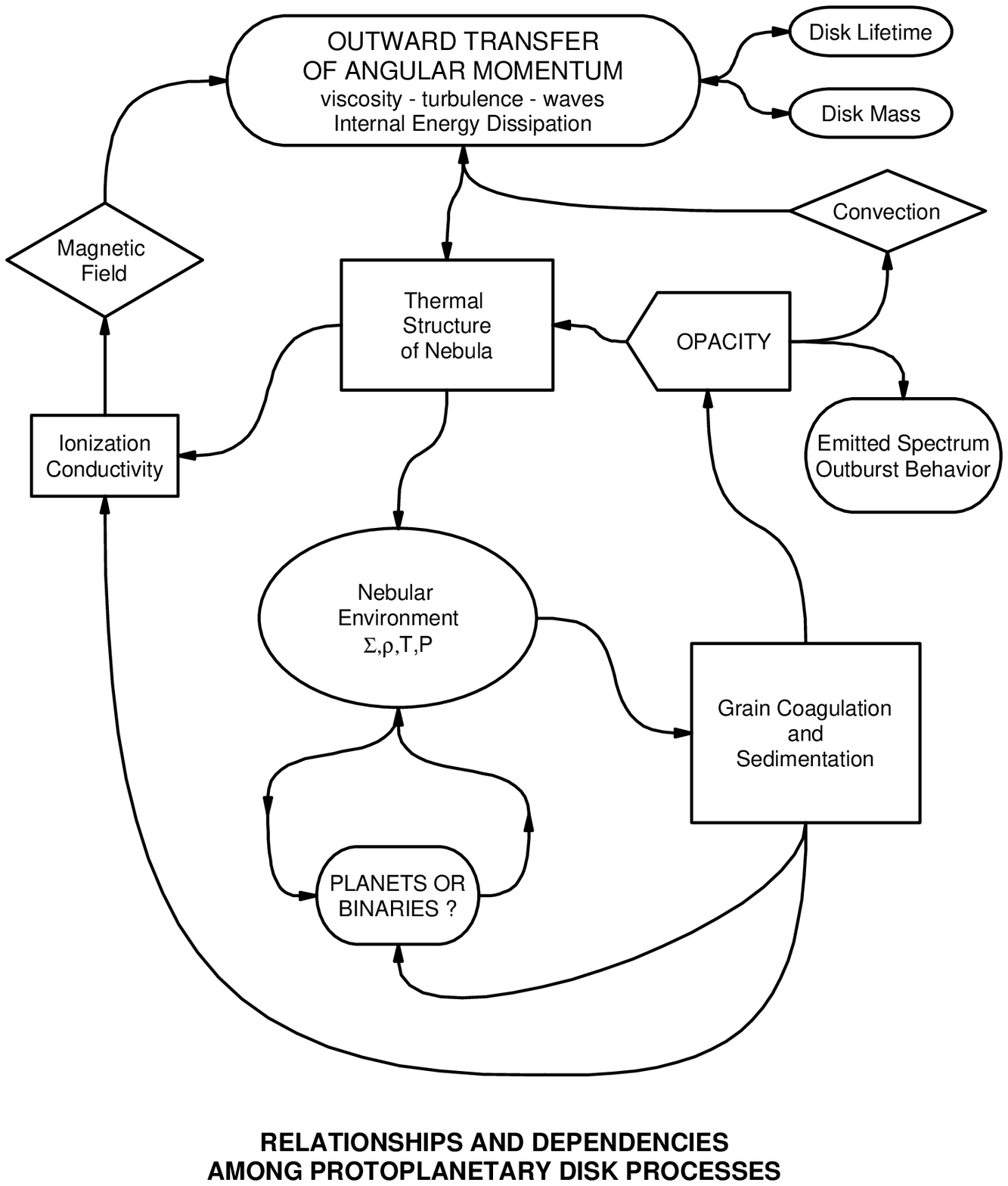}
\caption{Schematic diagram showing the relationships and feedbacks among
nebular processes, many of which are only poorly understood.  
Small modifications to the underlying physics
for any one process may have profound effects on the ultimate evolutionary 
outcome of the star/disk system.}
\label{nebuladeps}
\end{figure}

\section{From Grains to Planetesimals}
\label{sec:grainplanetesimal}

\subsection{Particle--Gas Dynamics}
\label{sec:particlegas}

The motion of solid particles in protoplanetary disks is affected by the
drag they experience as they move through the gaseous background of the nebula
(\refcite{whipple}; \refcite{adachi}; \refcite{stu77drag}).
We will consider the solids as spheres of radius $a$ having internal mass density $\delta$
(to be distinguished from the ambient gas density $\rho$) and mass $m= 4\pi \delta a^3/3$.
The form of the gas drag force, $\mathbf{F}_{\rm D}$, depends on the size of the particle,
its speed through the gas, and local gas properties.  
The Epstein drag law, which is appropriate for particles with
sizes smaller than the mean free path in the gas, is
\begin{equation}
\label{epstein}
{\mathbf F}_{\rm D} = - \frac{4}{3}\, \pi a^2 \rho \, c \, \mathbf{v},
\end{equation}
where $c$ is the gas sound speed and $\mathbf{v}$ is the velocity of the particle relative
to the gas.  In most regions of the solar nebula, this law is valid for particles with
sizes smaller than a few centimeters (see the references above for a complete list of 
the different drag law regimes); for simplicity of presentation in this review, I will adopt 
equation~(\ref{epstein}) as the principal drag formula.  The {\em stopping time} for a particle
is defined to be the time needed for drag to dissipate its momentum, $\tstop \equiv m v/F_{\rm D}$.
Substituting equation~(\ref{epstein}) into this definition and eliminating the volume
gas density in favor of the surface density from (\ref{surden}),
we find the relation
\begin{equation}
\label{stopt}
\Omega \tstop = \frac{\delta \, a}{\Sigma}.
\end{equation}
The dimensionless quantity $\Omega \tstop$ is the ratio of the
particle stopping time to the local orbital time in the disk and is a key determinant
of the dynamical behavior of the particle.  In the limit
$\Omega\tstop\ll 1$,
the particle, considered ``small'', is
very strongly coupled to the gas.  Drag will force it to move at the local gas velocity,
which we saw in \S\ref{sec:diskevol} to be sub-Keplerian.
In the opposite limit, $\Omega\tstop\gg1$, the particle, considered ``large'',
is only weakly perturbed by gas drag and orbits at the Keplerian speed.  

Let us consider the dynamics in more detail.  At each radius, gas drag forces
small particles to orbit at the local gas speed, which is 
smaller than Keplerian.  The net effective gravity (stellar plus centrifugal)
acting on the particle points toward the central star, and the particle drifts 
inward at the terminal speed.  This radial velocity is easily shown to be
\begin{equation}
\label{drift:small}
v_r = - 2 \,\Omega\tstop \, \Delta v,\ \ \ \ \ \ (\Omega\tstop\ll 1)
\end{equation}
where $\Delta v$ is the difference between the Kepler and gas speeds
defined in equation~(\ref{deltav}).  Large particles move at the Kepler velocity
and experience a headwind as they move through the more slowly moving gas.  The gas
drag torque causes the orbit of the particle to decay with radial velocity
\begin{equation}
\label{drift:big}
v_r = - \frac{2 \Delta v}{\Omega \tstop}. \ \ \ \ \ \ \ \ (\Omega\tstop\gg 1)
\end{equation}
Detailed calculations by \citeauthor{stu77drag} 
(\citeyear{stu77drag}; see Fig.~\ref{gasdrag}) verify these limiting behaviors
and show that the maximum inward drift speed is $\approx \Delta v$ and is achieved
by particles with sizes that satisfy $\Omega\tstop \approx 1$.  There are several points
to note about this inward drift.  First, it is size dependent leading to differential
speeds among particles with different radius,
which can enhance the rate of collisions.  Also, the relative speed between bodies
entrained in the flow can easily exceed their escape velocities (see Fig.~\ref{gasdrag}).
Second, the affect is largest for particles having radii for which $\Omega \tstop \approx 1$.
Using equation~(\ref{stopt}), we find that the size of these particles is 
$a \approx \Sigma/\delta$, which is of order 1~m at the position of the Earth.
Third, although the magnitude of the drift speed is highly subsonic, particles satisfying
$\Omega\tstop\sim 1$ decay into the central star in a time very much less than the evolutionary
time of the disk.  The drift time for these particles is 
$t_{\rm drift} = r/v_r = \Omega^{-1} \left(r/H\right)^2 \approx \alpha t_{\rm vis}
\approx 100 (r/{\rm AU})^{3/2}$
years.  Of course such particles are likely to collide with other solids on their way
in and through shattering or growth move to size regimes with smaller drift rates.
Nonetheless, gas drag-induced decay is a potentially large sink for the solid component of the
disk.

\begin{figure}[htb]
\epsfxsize=\hsize
\epsfbox{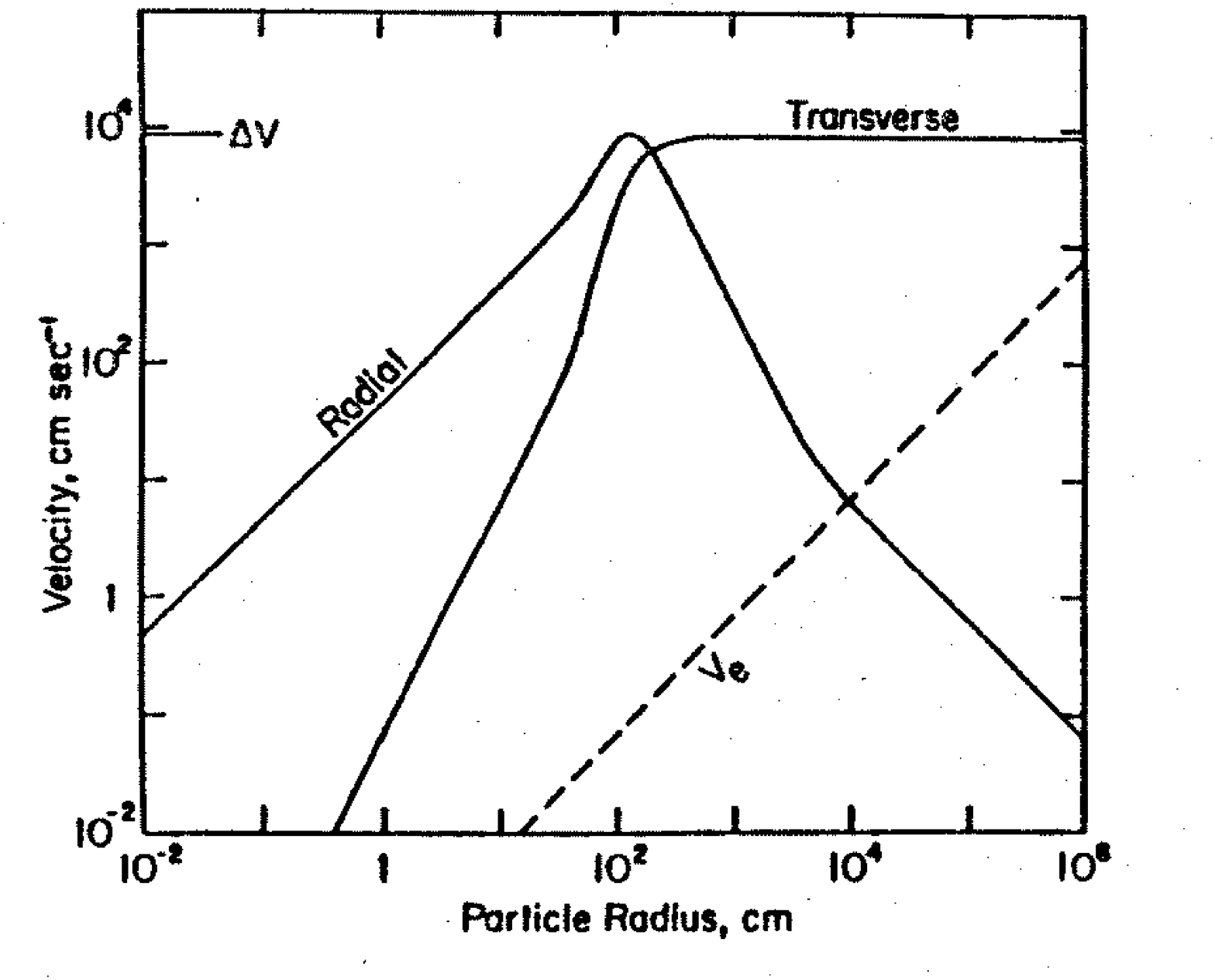}
\caption{Radial and transverse velocity components, relative to the gas, as a function
of particle radius.  The difference between the Kepler speed and the gas speed is
labeled $\Delta V$.
Small particles have transverse speeds equal to the gas speed,
while large particles move at the Kepler speed, {\em i.e.}, have transverse speeds relative
to the gas of $\Delta V$.  The maximum inward drift speed is $\approx \Delta V$ and is achieved
by particles with $\Omega\tstop \approx 1$.  The changes in slope are transitions between
drag regimes, and $V_{\rm e}$ is the escape speed from the particle.  From
\tcite{stu77drag}.}
\label{gasdrag}
\end{figure}

\subsection{Midplane Settling and Growth}
\label{sec:settle}

As discussed earlier, the formation of solid bodies in the midplane of the nebula begins with
agglomeration and sedimentation of smaller grains initially suspended throughout the nebula.
Small particles settle in the vertical gravitational field ($g_{\rm z} = (z/r) GM_\ast/r^2
=\Omega^2 z$) at the terminal velocity.  Using the Epstein law~(\ref{epstein}), the
settling speed and the time to reach the midplane are ($\Omega\tstop \ll 1$)
\begin{eqnarray}
  \label{vsettle}
v_{\rm z}  & =  & -(\Omega \tstop) \; \Omega z, \\
\label{tsettle}
t_{\rm settle} & = & \frac{z}{\vert v_{\rm z}\vert } = 
\frac{1}{\Omega}\, \left(\frac{1}{\Omega \tstop}\right) = \frac{1}{\Omega}\,
\left(\frac{\Sigma}{\delta\, a}\right).
\end{eqnarray}
The last equation shows that the settling time 
is always much longer than
the rotation period for small grains ($\Omega\tstop \ll 1$) and
that small grains
settle more slowly than larger grains ($t_{\rm settle} \propto 1/a$).
For typical solar nebula conditions
the settling time for micron-sized grains is about $10^6$ years,
comparable to the lifetimes of protoplanetary disks.
Without collisions to increase grain size (leading to more rapid settling),
small grains, which are the dominant contributor to the opacity, will 
remain suspended in the disk, which  will remain optically thick.

\tcite{safronov} was one of the first to realize that collisions
not only make particles grow but also rapidly decrease the settling time.
Equations~(\ref{vsettle}) and (\ref{stopt}) show that the
settling speed is proportional to the size of the particle.
Larger grains will sink towards the midplane faster and will sweep
up smaller grains in their path.  This causes them to grow larger and to sink
even more rapidly.  In this manner, the dust component of the disk
will therefore ``rainout'' and form a thin layer of much larger bodies in the midplane
of the nebula.  We can estimate the size of these bodies by using mass conservation.
The total surface density of solids in the disk is $\zeta \Sigma$, where $\zeta$
is the dust-to-gas ratio introduced above.  A body that
reaches the midplane with final radius $a$ 
will have swept up a column of dust having a mass $\pstick \cdot 
\zeta \Sigma \cdot \pi (a/2)^2$, where we have taken its average radius on
the descent to be $a/2$ and have introduced the probability
of sticking in a collision, $\pstick$.  Equating this to its mass $m=4\pi\delta a^3/3$
yields a size
\begin{equation}
\label{deltaa}
a = \frac{3 \, \pstick \, \zeta \, \Sigma}{16 \, \delta},
\end{equation}
which is about 1~cm at the Earth in the primordial solar nebula (for $\pstick=1$).
The time for solids to sediment to the midplane can be estimated by inserting
this value for $a$ into equation~(\ref{tsettle}); more accurate integrations
yield (\refcite{nakagawa})
\begin{equation}
\label{tsediment}
t_{\rm sediment} \approx \frac{100}{ \Omega} \; \frac{1}{\pstick \, \zeta},
\end{equation}
which is only a few thousand orbital periods  at any
point in the disk (again for $\pstick = 1$).

There have been several detailed numerical simulations of the
collisional growth and sedimentation of dust in the solar nebula
(\refcite{studust}; \refcite{naknakhay}; \refcite{mizunodust}; \refcite{stucuzzi}).
A major difficulty with simulating particle growth is our inadequate
knowledge of the physical inputs.  The sticking probability is a 
complicated and currently only poorly 
understood function that depends on the impact velocity and the 
size, shape and internal strength of the impactors, 
but it is usually simply taken to be unity for all collisions 
in the simulation.  
Realistic nebular particles may have
low internal strengths and may  shatter in high speed impacts.
Some experimental work has been performed on this important issue.
\tcite{blum} did not find {\em any} sticking in collisions between millimeter-sized
dust aggregates at speeds $\lta 1$~m/s and found fragmentation at speeds above this.
More recently, this group (\refcite{wurm}) has investigated lower speed
($\lta 1$~cm/s) collisions between small, porous, fractal grains with
sizes $\sim 10\; \mu$m and found perfect sticking ($\pstick =1$).  
In the outer, cooler regions of the nebula volatile ices can form frosts 
on particle surfaces, which may enhance the sticking probability
(\refcite{bridges}; \refcite{supulver}).

Turbulence in the gas can have profound affects on the
evolutionary simulation.  
A given sized particle will respond to the 
fluctuating turbulent velocity field according to whether its stopping time
is longer or shorter than the local eddy turnover time.  
Stirring by turbulence can inhibit sedimentation
of particles that have sizes small enough that they are strongly coupled to
and co-move with the turbulent eddies.  The turbulent velocity field
can also greatly increase the relative velocity between particles with
different sizes (\refcite{markiewicz}), which can increase the collision
rate but also lead to collisions where the bodies shatter
(\refcite{studust}).  
Given our poor understanding of 
turbulence in protoplanetary disks, its effects on grain growth (and vice versa)
and grain sedimentation are presently far from clear. 
Even with these uncertainties and although the individual details
vary, the different numerical simulations nonetheless yield
midplane particle sizes and sedimentation timescales that are not too dissimilar from
the simple estimates above (eqs.~[\ref{deltaa}] \& [\ref{tsediment}]).

\subsubsection{Radial Distribution of Solids in Protoplanetary Disks}

The sedimentation of dust grains produces a thin layer of solids in the
nebular midplane.  The surface density of this distribution is often
taken to be just the gas-to-dust ratio times the local gas surface density,
$\zeta\Sigma$.  To test whether this is a valid approximation and to see
whether the solid surface density distribution of the \mmsn\ is reproduced, 
I have calculated simplified models of the midplane accumulation
of solids using the nebular evolutionary models of \tcite{rudpoll}.
The models follow track the condensed and vapor phases of refractory rocky and volatile
icy matter, with ices condensing only in nebular regions with temperatures below
170~K.  Early in the evolution, the nebula is too hot for ices to condense anywhere,
although rocky matter can condense everywhere outside about 0.1~AU.
As the nebula evolves and cools, the radius where ices can condense,
{\em i.e.}, the $T=170$~K boundary, moves inward and icy solids sediment.
The treatment of grain growth and sedimentation is phenomenologically
modeled by allowing the
condensed solids to rainout on a timescale given by
$\tau_0/\zeta\Omega$  ({\em c.f.} eq.~[\ref{tsediment}]),
where $\tau_0$ is an adjustable constant.
This model is nonlinear
because as the solids sediment to the midplane, the  dust-to-gas ratio $\zeta$
decreases, and the  sedimentation time lengthens.  
Figure~\ref{sprmmsn} shows the results of two evolutionary scenarios: the first with
rapid sedimentation, $\tau_0 = 600$, and the second with much slower sedimentation,
$\tau_0=6\times10^4$.  Both models started from the same initial conditions.
Significantly more solids are deposited in the midplane
in the rapid scenario, but at the time shown in the plot ($10^5$ years) 
the nebula is warm enough
that ices only condense outside 6~AU.  The results of the slow scenario give a 
surface density distribution comparable to the \mmsn\ after $\sim 10^6$ years,
but with a more shallow slope (and more solid matter) in the outer cool regions.

It seems clear from these idealized models that very different distributions of
midplane solid matter are possible, from the same initial mass reservoir, if the 
sedimentation physics (primarily the average sticking probability) is different.
The solid distributions shown in Figure~\ref{sprmmsn} do {\em not}
bear a simple constant multiplicative relationship ($\zeta\Sigma$) to the 
instantaneous gas surface  density distribution.  
The reason for this is the solids at any given radius 
were deposited over time and reflect accumulation from 
different nebular conditions, with primarily rocky material 
sedimenting early on when the temperature was high followed by icy sedimentation
when the nebula was cooler.  Note also that the locus at which ice can condense
is different in the two models; in general, the midplane distance at which
$T=170$~K (the ice condensation front) is a function of time.

\begin{figure}[htb]
\epsfxsize=\hsize
\epsfbox{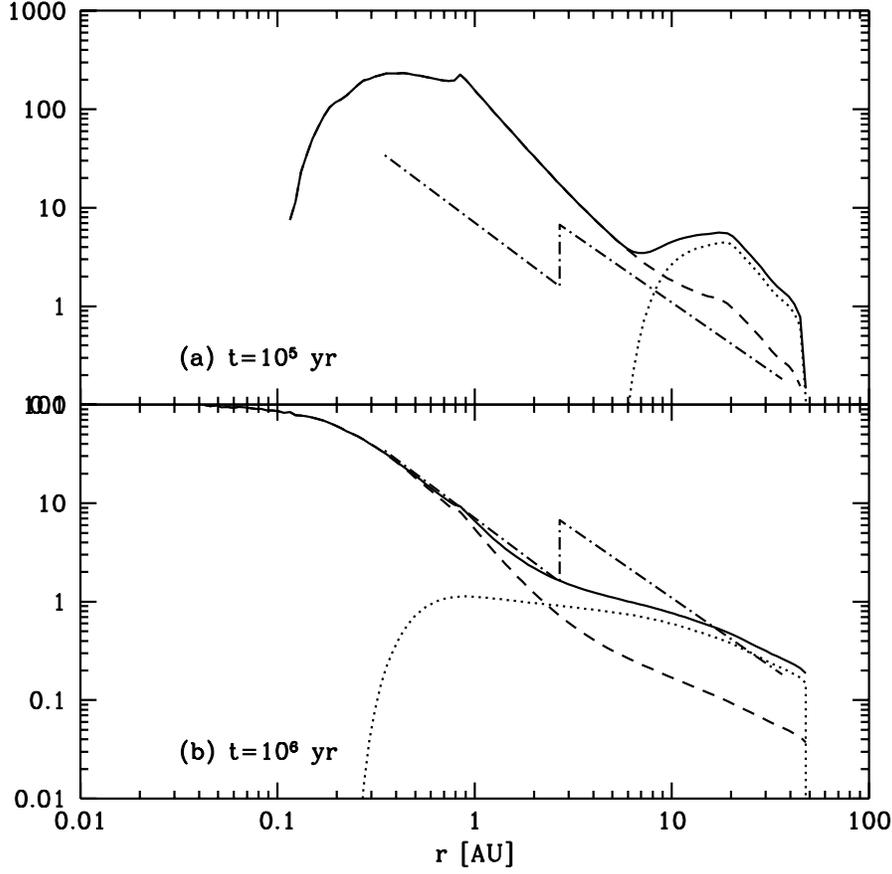}
\caption{Radial distribution of solids in a model of grain growth and
sedimentation (\refcite{ruden98}).  Surface density (in g/cm$^2$) is
plotted versus distance in AU.  The surface densities of icy (dotted),
rocky (dashed), and rock $+$ ice (solid) matter are plotted.
The \mmsn\ surface density distribution is shown for comparison (dot-dashed).
Panel (a) [(b)] is for the rapid [slow] accumulation scenario, with the evolution
times indicated.  Note the difference in vertical scale between the panels.
See text for further details.}
\label{sprmmsn}
\end{figure}

\subsection{Planetesimal Formation}
\label{sec:planetesimal}

Planetesimals are bodies large enough to move on Keplerian orbits largely
unaffected by gas drag forces over the age of the disk.  For typical nebular
conditions, bodies larger than about 1~km satisfy this constraint.  
A key unanswered question is how the smaller sized bodies ($\sim 1~\mbox{cm}$;
eq.~[\ref{deltaa}]) produced in the sedimentation process assemble into planetesimals.
The most elegant and efficient mechanism proposed is that the thin layer of solids
produced by sedimentation becomes gravitationally unstable and rapidly forms
fragments that collapse to produce planetesimals (\refcite{GW}).  
They showed the critical fragmentation lengthscale, $\lambda_c$, is
\begin{equation}
\label{GWlength}
\lambda_c \approx 4 \pi r \; \left(\frac{\pi \zeta \Sigma r^2}{M_\ast}\right),
\end{equation}
and the typical fragment mass, $M_{\rm GW}$ is
\begin{equation}
\label{GWmass}
M_{\rm GW} = \zeta \Sigma \lambda_c^2 
= 16 \pi M_\ast \; \left(\frac{\pi \zeta \Sigma r^2}{M_\ast}\right)^3.
\end{equation}
In these equations, it is the surface density of {\em solids}, $\zeta \Sigma$, that
is important, and the ratio in parenthesis is roughly the mass of the solid disk
to the protostellar mass.
Using nebular parameters from the \mmsn,
\tcite{GW} showed that kilometer-sized bodies can be
produced by this gravitational collapse process in a few thousand years.

The Goldreich-Ward instability occurs only in very quiescent dust disks.  If the random
velocities of the particles composing the dust disk are too large (or equivalently
if the dust layer is too thick)\footnote{The factor in parentheses in equations~(\ref{GWlength})
\& (\ref{GWmass}) is the {maximum} value of the aspect ratio, $H/r$, the layer
can have to be gravitationally unstable.} 
then the instability is quenched.  The critical
random speed is $\pi G \zeta \Sigma/\Omega$, which is only about 10~cm/s.
In the thin dust disk the solids rotate at nearly the Keplerian velocity while
the gas above the layer rotates at a sub-Keplerian rate (eq.~[\ref{deltav}]).
It is thought that turbulence generated by this shear layer will stir up the 
random velocity in the dust disk and prevent the Goldreich-Ward instability (\refcite{stucuzzi}).
Detailed hydrodynamical models of the particle-gas interaction have been
made by \tcite{cuzzidust} who find that it is unlikely the dust layer can achieve such
low random speeds.

If the Goldreich-Ward mechanism does not proceed in protostellar disks, then it
is likely that planetesimals are produced in the thin dust layer by inelastic binary
collisions.  Recall that gas drag causes radially inward flow of solid particles
(eqs.~[\ref{drift:small}] \& [\ref{drift:big}]).  In the dense particle layer,
solids dominate the local density and the drift speed is reduced
(\refcite{cuzzidust}).  Since larger particles have lower drift speeds
(eq.~[\ref{drift:big}]), \tcite{cuzzidust} have argued that planetesimals can grow
in this layer as the smaller solids drift past them and are collisionally accreted.
They estimate that growth to 100~km sizes can be achieved in $\lta 10^6$ years.

\subsection{Prospects for Observing the Early Planet Formation Epoch}
\label{sec:observable}

It would be satisfying if theoretical models gave an unambiguous prediction
for the observational signature of grain growth in young protoplanetary disks,
for example, the existence of a ``hole'' in the infrared spectrum 
(such as the one inferred observationally in HR~4796 by \refcite{koerner}).
Unfortunately, this is presently not the case.  
In order to ``see'' forming 
planets or other structures in the midplane of the
disk, the optical depth to the nebular midplane must be small.
The disk optical depth is equal to the gas surface mass density times 
the opacity.  The midplane regions can remain cloaked to the eyes of  observers
if {\em either} is sufficiently large.
The primary contributors to the opacity are small grains, but its value
even for the standard 
interstellar grain-size distribution is subject to debate (\refcite{becksargIII};
see the review chapter by Beckwith); grain size evolution only complicates the 
picture (\refcite{miyake}).

As noted in \S\ref{sec:settle}, micron-sized grains take times comparable to
the age of disks to settle in quiescent nebulae. 
Any small grains left suspended in the disk after the bulk of the solid mass
has sedimented to the midplane will remain suspended in the disk as long as it is 
quiescent.  The effects of nebular turbulence on a suspended small grain population
are harder to discern.
By itself, turbulence causes small particles to remain stirred up and
suspended in the nebula.  However, turbulence also mixes these particles into
the midplane region where they may collide and stick to larger bodies
(\refcite{stucuzzi}).  The net result of these two competing processes 
on the small grain population is unknown.  Also uncertain is how the turbulent
properties change as the disk optical depth changes, {\em i.e.} whether the
disk will remain turbulent as grain growth proceeds.

Current theoretical models cannot reliably predict how much of the 
initial solid mass will remain leftover as small particles, 
from processes such as particle shattering and erosion.
The nebular models of \citeauthor{rudlin}~(\citeyear{rudlin}, Fig.~\ref{nebmodels})
and \tcite{rudpoll} remain 
optically thick through mid-infrared wavelengths at all disk radii as long
as $\gta 1$~\% of the initial
grain mass remains suspended.  The \mmsn\  (Fig.~\ref{mmsnfig})
will also remain optically thick out to $\lta 10$~AU under the same conditions.
Finding observable signatures may require waiting until the disk gas is
cleared (probably taking with it the small grain population) or searching for
planet formation signatures in long wavelength regimes that remain optically thin.

\subsection{Future Research Questions}

\begin{itemize}
\item What are the appropriate conditions in protostellar disks at the
inception of the planet building epoch?
	\begin{itemize}
		\item Do they resemble the \mmsn?
		\item Do different initial conditions lead to diverse planetary systems?
	\end{itemize}
\item What is the sticking probability in  grain-grain collisions?
	\begin{itemize}
		\item Do simulations using $\pstick=1$ adequately model
			grain growth?
	\end{itemize}
\item Is planet formation inefficient?
	\begin{itemize}
	\item How much solid mass is lost via gas-drag induced radial flow?
	\end{itemize}
\item How are planetesimals formed?
	\begin{itemize}
	\item Via the Goldreich-Ward instability or by collisions?
	\end{itemize}
\item What are observable signatures of planet formation?
	\begin{itemize}
	\item Holes, or gaps, or warps?
	\end{itemize}
\end{itemize}

\section{From Planetesimal to Planet}

Planets are formed from the midplane planetesimal distribution by inelastic 
collisions.  The key feature of the collisional evolution is that the
cross-section for pairwise collisions can be increased above the geometrical
value by gravitational attraction between the colliders (\refcite{safronov}).  
This result can be shown quite easily by considering energy and angular momentum
conservation in the collision between two particles with masses $m_1$ and
$m_2$ and radii $a_1$ and $a_2$.  In the center-of-mass system, the constant orbital
energy and angular momentum are
\begin{eqnarray}
\label{energy:E}
E & = & \frac{1}{2} \mu v^2 - \frac{G m_1 m_2}{r} =  \frac{1}{2} \, \mu V^2,\\
\label{energy:J}
J & = & \mu \left\vert {\mathbf r} \times {\mathbf v} \right\vert = \mu r v \sin\theta =
\mu b V,
\end{eqnarray}
where $\mu=m_1 m_2/(m_1+m_2)$ is the reduced mass.
In the rightmost equalities I
have evaluated the constants $E$ and $J$ when the particles are at large separation
where $b$ is the impact parameter and $V$ is the relative 
collision velocity.
Without gravity the maximum impact parameter that results in a collision is 
the sum of the radii $a_1 + a_2$.
With gravity, a grazing collision results when 
the minimum
separation is $r=a_1+a_2$ and the velocity is purely tangential, $\theta=\pi/2$.
Substituting these two conditions into equation~(\ref{energy:J}) gives the speed at
impact as $v=bV/(a_1+a_2)$, which can be substituted into equation~(\ref{energy:E}) to give
the gravitationally enhanced cross-section for collision
\begin{equation}
\label{xsection}
\pi b^2 = \pi \left(a_1+a_2\right)^2 \; \left[ 1 + \frac{v_{\rm e}^2}{V^2}\right],
\end{equation}
where the mutual escape speed, $v_{\rm e}$, is
\begin{equation}
\label{vescape}
v_{\rm e} = \sqrt{\frac{2 G \left(m_1 + m_2\right)}{a_1 + a_2}}.
\end{equation}
The term in brackets in equation~(\ref{xsection}) is the gravitational focusing factor
and can greatly increase the cross-section above geometrical if the relative 
collision speed, $V$, is much smaller than the escape speed, $v_{\rm e}$.  
Also note that if the focusing factor is large the cross-section is proportional to the fourth
power of the particle radius (because $v_{\rm e}^2 \propto a^2$) rather than the second.
In honor of Safronov's pioneering contributions to the study of collisional
planet growth, the gravitational focusing factor is often written as $1 + 2\theta$, where
$\theta \equiv v_{\rm e}^2/2 V^2$, is called the Safronov number.
Detailed three-body 
numerical orbit integrations confirm that the cross-section (\ref{xsection})
is a good approximation (\refcite{ida}; \refcite{greenz1}, \citeyear{greenz2}).

The equation for the collisional growth of a planet with mass $\mplanet
= 4 \pi \delta a^3/3$ (recall $\delta$ is the internal density of the planet)
accreting matter from a background ``swarm'' of planetesimals that move
with respect to it at a relative speed $V$ is (using the cross-section [\ref{xsection}]
and neglecting the size of the planetesimals)
\begin{equation}
\label{planetmass}
\frac{d \mplanet}{d t} = \rho_{\rm s} \cdot V \cdot \pi a^2 \left[ 1 + \frac{v_e^2}{V^2}\right],
\end{equation}
where $\rho_{\rm s}$ is the volume density of the planetesimal swarm.
We can rewrite this as
an equation for the rate of change of the planetary radius
\begin{equation}
\label{planetradius}
\frac{ d a}{d t} = \frac{1}{8} \, \left(\frac{\Omega \Sigma_{\rm s}}{\delta}\right)
\, \left[1 + \frac{v_e^2}{V^2} \right],
\end{equation}
where $\Sigma_{\rm s} = 2 \rho_{\rm s} V/\Omega$ is the surface density of the 
planetesimal swarm ({\em c.f.} eq.~[\ref{surden}]).

\subsection{Orderly Growth}
\label{sec:orderly}

The parameter that plays the most crucial role in determining 
the type of planetary growth is the Safronov number, {\em i.e.},
the magnitude of the planetary escape speed relative to 
the planetesimal velocity dispersion.
Using a variety of analytical approaches, 
\tcite{safronov} argued that as planetesimals accumulated, gravitational
scattering off the largest member of the swarm would cause the velocity
dispersion to keep pace with the escape speed from that member, regulating
$\theta$ to be near unity.  In this scenario planets grow in an orderly fashion
with most of the mass in the large bodies.  
From equation~(\ref{planetradius}), the accumulation time is
\begin{eqnarray}
\label{orderly}
t_{\rm orderly} & = & \frac{a}{da/dt} \approx \frac{1}{\Omega} \, 
\left(\frac{\delta \, a}{\Sigma_{\rm s}}\right),\ \ \ \ \ (\theta \sim 1)\\
& \approx & 10^8 \, \left(\frac{r}{{\rm AU}} \right)^3 \ \mbox{years}.\nonumber
\end{eqnarray}
The growth time is determined by both the Kepler clock and the solid surface density.
In the last line I have evaluated the growth time for Earth-sized bodies and taken
$\Sigma_{\rm s}$ from the \mmsn\ distribution
(Fig.~\ref{mmsnfig}).  These timescales are quite
long, far greater than the lifetime of nebular disks, with the orderly growth
process being unable to produce the cores of the outer giant planets in less than the
age of the universe!

\subsection{Runaway Growth}
\label{sec:runaway}

The key to forming planets more rapidly is keeping the velocity dispersion
small enough so that the gravitational focusing factor  becomes very large ($\theta \gg 1$).  
In this case, the most massive planetesimal in the swarm will have the largest 
collisional cross 
section ($\propto a^4$), will grow more rapidly than any other body and
will separate itself from the remainder of the swarm.  
Numerical models by \tcite{greenberg} were the first to find this ``runaway'' path
to planetary growth; subsequent calculations have confirmed these models
(\refcite{wetherstew},~\citeyear{wetherstew2}; \refcite{ohtsuki}; \refcite{aarseth}).
The basic mechanism behind runaway growth was elucidated by 
\tcite{wetherstew} \& (\citeyear{wetherstew2})
Planetesimal velocities are
determined by a balance among 1) stirring by gravitational scattering,
2) stirring by inelastic collisions, 3) damping due to energy dissipation in
inelastic collisions, 4) damping due to gas drag, and 5) energy transfer 
from large to small bodies via dynamical friction (\refcite{lissstu}).
The most critical factor is dynamical friction, which tends to lead to
an equipartition of kinetic energy among the bodies (\refcite{binney}),
with the largest bodies having the smallest random velocities and thus the
largest cross-sections.  Figure~\ref{WSrunaway} illustrates the runaway
process in which the largest body grows the fastest as long
as the velocity dispersion remains low.

\begin{figure}[htb]
\epsfxsize=\hsize
\epsfbox{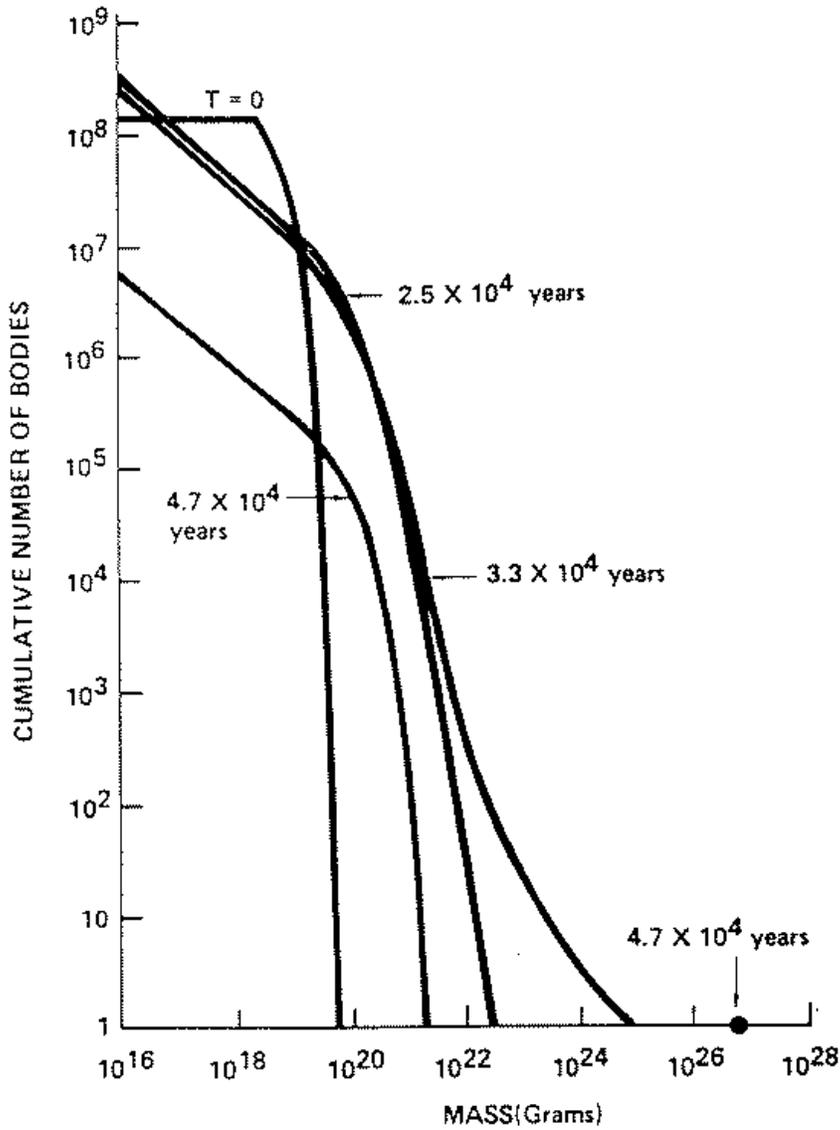}

\caption{The evolution of the mass distribution of a planetesimal swarm
between 0.99 and 1.01~AU illustrating the rapid runaway growth of the largest bodies.
The most massive body becomes detached from the remainder of the swarm 
in less than $5\times 10^4$~years.  See \tcite{wetherstew}.}
\label{WSrunaway}
\end{figure}

Runaway growth can proceed until all the available material within the ``accretion
zone'' (sometimes called the ``feeding zone'') of the
planet is consumed (\refcite{wether90}; \refcite{lisrunaway}, \citeyear{lisreview}).
The accretion zone is an annulus over which the planet can exert its gravitational
influence to perturb nearby planetesimals into crossing orbits.  
The most common measure of the gravitational range of a planet of mass $\mplanet$
a distance $r$ away from a primary body of mass $M_\ast$
is the Hill sphere (or Roche lobe radius) defined to be
\begin{equation}
\label{hillsphere}
r_{\rm H} = r \; \left(\frac{\mplanet}{3M_\ast}\right)^{1/3}.
\end{equation}
The origin of the cube root mass dependence comes from finding the 
region over which  the 
gravity of the planet, $G\mplanet/r_{\rm H}^2$,
dominates the tidal force of the primary object, $(GM_\ast/r^2) (r_{\rm H}/r)$.
The radial
size of the accretion zone has been estimated to be a numerical factor $B \approx 4$
times larger than the Hill sphere of the planet (\refcite{lisreview}).
If the planet accretes all the available solid mass (with surface density $\Sigma_{\rm s}$)
in an accretion zone having width $B\, r_{\rm H}$ on either side, its final ``isolation'' mass
will be
\begin{eqnarray}
\label{runawaymass}
\mplanet &=& 2 \pi r \cdot 2B r_{\rm H} \cdot \Sigma_{\rm s} = 4\pi B r^2 \Sigma_{\rm s} 
\left(\frac{\mplanet}{3M_\ast}\right)^{1/3}\nonumber \\
& = & \frac{\left(4 \pi B r^2 \Sigma_{\rm s}\right)^{3/2}}{\left(3 M_\ast\right)^{1/2}}.
\end{eqnarray}
The radial spacing of planets that have each become isolated is $\approx 2 B r_{\rm H}$.
For the \mmsn, we find an isolation mass of 0.05$\earth$ at 1~AU and 1.4$\earth$ 
at Jupiter's distance.

\begin{figure}[htb]
\epsfxsize=\hsize
\epsfbox{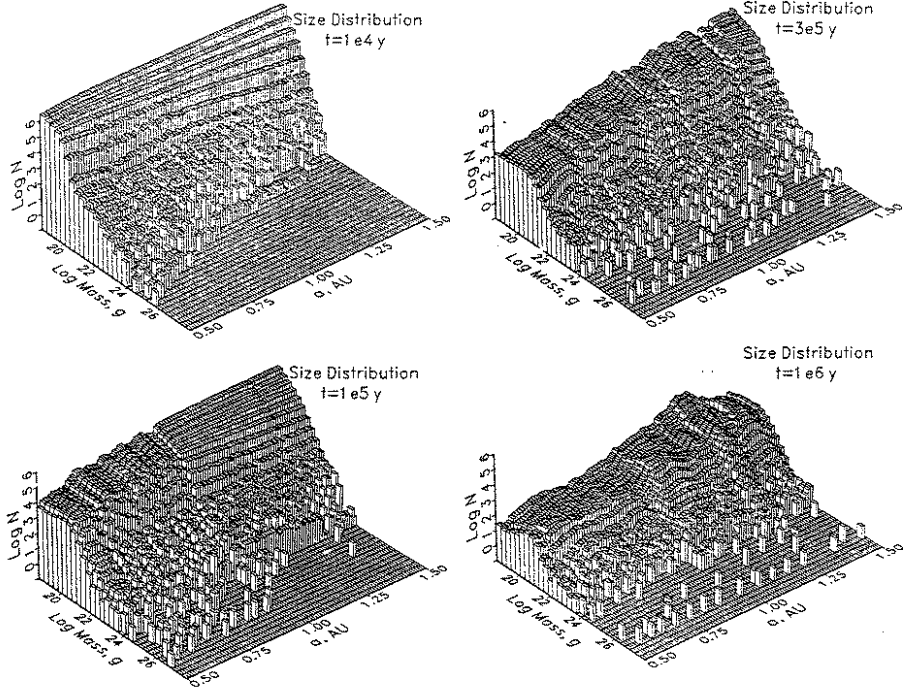}
\caption{Accretional evolution of a planetesimal swarm
near 1~AU.  The number of planetesimals 
are plotted per logarithmic interval in mass as a function of distance from the Sun.
The evolution is clearly more rapid at 0.5~AU than at 1.5~AU, but the
final mass of the protoplanets ($\sim 10^{27}$~g) is not dependent on
distance from the Sun.  Runaway growth is evident as the largest bodies separate
themselves from the rest of the planetesimal population.  After $10^6$ years there
are more bodies with lower mass than in the terrestrial region of the Solar System.
From \tcite{stuplanet}.}
\label{sturunaway}
\end{figure}

A recent detailed simulation of the accumulation of planetesimals 
in the terrestrial planet region has been performed by \tcite{stuplanet}
that includes the processes mentioned above that control the velocity 
dispersion plus orbital perturbations caused by distant planetary bodies.
The results of a simulation of the accumulation of $2\;\earth$ of 
initially
$4.8\times10^{18}$~g (15~km)
planetesimals spread between 0.5 and 1.5~AU is illustrated in Figure~\ref{sturunaway}.
These authors verified the critical role dynamical friction plays in producing
rapid runaway growth.  The first panel in Figure~\ref{sturunaway} ($t=10^4$~years)
clearly shows the evolution to larger masses occurs more rapidly at smaller radii
where the Kepler frequency is larger.
The other panels show that more massive bodies runaway and detach themselves from the
distribution of lower mass planetesimals.
Runaway growth produces bodies with masses $\sim 10^{26}$~g in $\lta 10^5$~years,
but the growth rate slows as the velocity dispersion of the population of lower mass
bodies increases, and the runaway stalls.  By $10^6$ years, about a dozen 
bodies with masses $\approx 10^{27}$~g have been produced,
but they are lower in mass and more closely spaced than the
terrestrial planets.  The masses (and separations) found in the simulation
are about a factor of three larger than the 
analytic isolation mass in equation~(\ref{runawaymass}).

\subsection{Final Accumulation Stages}
\label{sec:finalacc}

The outcome of runaway growth is self-limiting for two reasons.
First, if the mass of the largest
bodies becomes $\sim 100$ times the median mass of the continuum bodies,
the velocity dispersion of the
lower mass planetesimals is pumped up by gravitational scattering, 
and the runaway slows (\refcite{idamak}; \refcite{stuplanet}).
Second, even if the velocity dispersion
remains small,
all the mass within the accretion zone will be consumed by the planet,
and it will become dynamically isolated with a mass given by 
equation~(\ref{runawaymass}).
The evolution following 
runaway is much longer term as gravitational encounters slowly perturb the bodies
into crossing orbits and violent impacts occur.  
The models described in \S\ref{sec:runaway}, which start from \mmsn-type 
conditions, do not yield masses and spacings
similar to those of the terrestrial planets, hence,
longer term collisional evolution is necessary to produce a final
model that is similar to the Solar System.
Figure~\ref{earthgrow} shows the results of six accumulation calculations by
\tcite{wether88} that started with 500 bodies each of mass $2\times10^{25}$~g.
The outcome of each simulation is a small number of planets after about $10^8$ years.
The differences in the calculations illustrate the stochastic nature of the collisional
accumulation.  The long accumulation timescales do not present any serious difficulties
for the terrestrial planets, which as noted in \S\ref{sec:over} could have completed
their evolution in a gas-free environment.  However the giant planet cores had to form in
less than the lifetime of the gas disk, which seems to require a rapid runaway growth
process occurring in a disk with solid mass larger than the \mmsn\ (\refcite{lisrunaway},
\citeyear{lisreview}; eq.~[\ref{runawaymass}]).

\begin{figure}[htb]
\epsfxsize=\hsize
\epsfbox{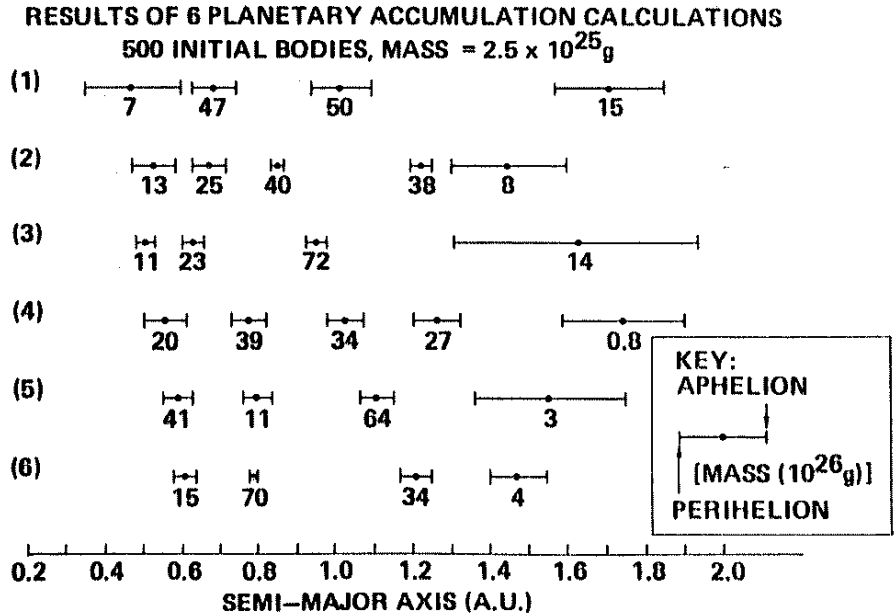}
\caption{Six calculations of terrestrial planet accumulation by \tcite{wether88}.
The final mass of each planet in units of $10^{26}$~g is indicated ($1 \earth = 60$
in these units).  The
semimajor axis, with a bar extending from perihelion to aphelion, is also shown.
From \tcite{lisreview}.}
\label{earthgrow}
\end{figure}

\subsection{Future Research Questions}

\begin{itemize}
\item Can the Earth and the cores of giant planets
be built directly from the runaway or is further long-term collisional
growth needed?
	\begin{itemize}
	\item How do you build the giant planet cores before the 
	gas is dispersed?
	\end{itemize}
\item How does the final planetary system depend on the initial
disk mass distribution?
	\begin{itemize}	
	\item Do higher mass disks produce {\em more} planets
	or  fewer, but higher-mass, planets?
	\end{itemize}
\item Can {\em rocky} planets with masses comparable to  Jupiter be built in protoplanetary disks?
\end{itemize}

\section{The Formation of Gas Giant Planets}
\label{sec:giants}

\subsection{The Core-Instability Scenario}

The current model for the formation of gas giant planets is that
they follow the path outlined above and begin as rocky cores.  \tcite{mizuno}
showed that a gas envelope gravitationally bound to a rocky core cannot
remain in hydrostatic equilibrium if the core mass exceeds a critical value $\sim 10 - 15 \; \earth$,
comparable to the core masses inferred for the giant planets (\refcite{hubbard};
\refcite{chabrier}).
Subsequent time-dependent calculations (\refcite{bodenheimer}; \refcite{pollack};
see top panel in  Fig.~\ref{coreinstability}) have shown 
the nebular gas is rapidly accreted by the rocky core once its mass is above critical, leading
to the conversion of a rocky planet into a gas giant planet.
Before the critical core mass is reached, the  radiated 
luminosity of the planet is supplied primarily
by the gravitational energy of the accreted planetesimals.  As the 
gravitationally bound gas envelope becomes more
massive, planetesimal accretion is unable to supply the luminosity needed to maintain 
the extended envelope and rapid (but not hydrodynamic) collapse of the envelope takes place, 
with the gravitational contraction of the envelope supplying the luminosity.
As the envelope contracts, surrounding nebular gas flows in, and the gas accretion rate rises
to become orders of magnitude larger than the solid accretion rate.
The evolutionary timescale shortens; the gas envelope of Jupiter is accreted in $\lta 10^5$~years
(\refcite{pollack}).  Detailed calculations show the value of the critical core mass ($\sim 10 -
30\;\earth$) is only moderately dependent on orbital radius and nebular properties and depends
most strongly on the planetesimal accretion rate (see \citeauthor{stevenson} [\citeyear{stevenson}]
for a simple analytical calculation of the critical core mass).

\begin{figure}[htb]
\epsfxsize=0.9\hsize
\epsfbox{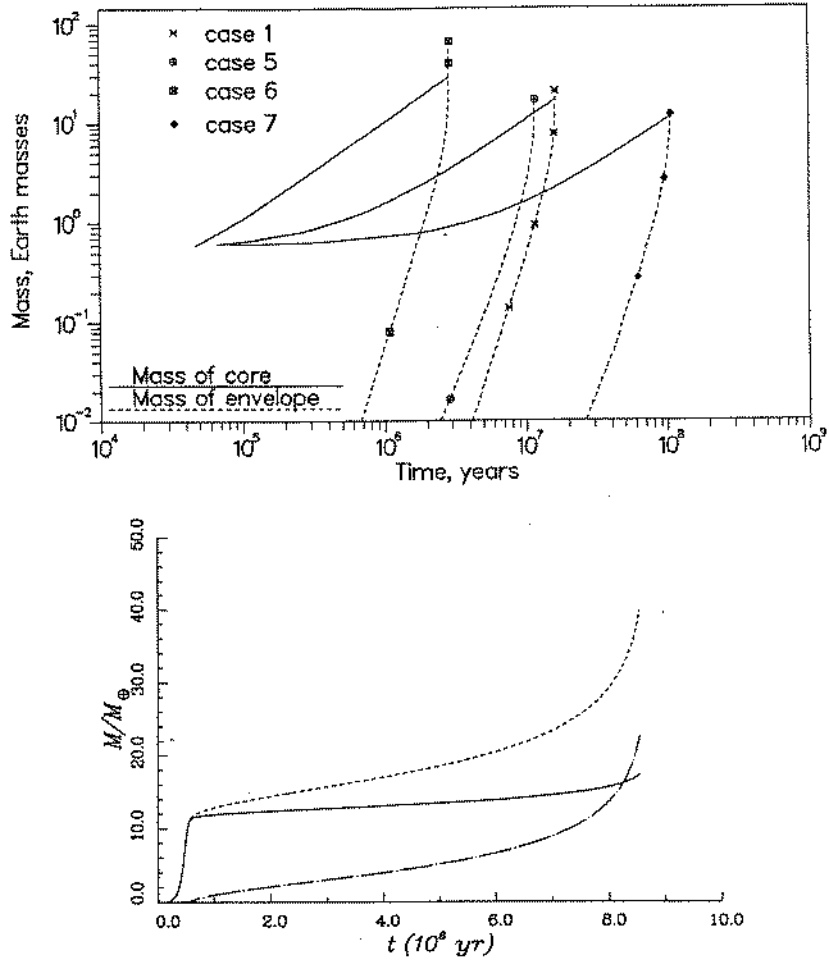}

\caption{ Core-Instability Model.  \textbf{Top panel} from
\tcite{bodenheimer} shows the core mass (solid) and envelope mass
(dotted) as a function of time.  Cases 6, 1 \& 5, and 7 have
constant planetesimal accretion rates of $10^{-5}$, $10^{-6}$, and
$10^{-7}~\earth$/yr, respectively.  Case 5 has grain opacity reduced
by 50 to reflect the decreased dust-to-gas ratio after solids have
sedimented from the gas. The envelope mass is very small until the
critical core mass is reached (defined to occur when $M_{\rm core} =
M_{\rm env}$).  At that point rapid gas accretion ensues.
\textbf{Bottom panel} from \tcite{pollack} shows the core mass
(solid), envelope mass (dot-dashed) and total mass (dotted) as a
function of time.  Rather than assuming constant planetesimal
accretion rates (as in the top panel), this calculation uses more
detailed and self-consistent planetesimal accretion rates.  The
initial core forms by runaway in less than $10^{6}$ years
after which the growth rate slows.  The planet becomes critical after 
$\approx 8\times 10^6$ years, when rapid gas accretion begins.}
\label{coreinstability}
\end{figure}

In the core-instability model, giant planets are composite bodies with the rocky cores forming
first and the gas envelopes accreting afterwards.  If this model is correct,
there is a logical simplicity to the formation
of planetary systems - all planets are born by the accumulation of solids; giant planets 
are higher mass cores that
happen to accrete large gas envelopes.  
The core-instability model is not without problems, however.  First, it is not obvious that
$\gta 10\earth$ solid cores can form before the nebular gas is dispersed.  Runaway growth to 
the critical mass requires solid densities about four times larger than in the \mmsn\ 
(see eq.~[\ref{runawaymass}]).  Forming the cores in $\sim 10^6$ years requires a large
net gravitational focusing factor ($\theta \gta 10^3$), which requires a very
cold population of accreting planetesimals.  Even if such cores can be formed at
the orbit of Jupiter, it is not clear they can be formed so quickly at 
the orbits of Uranus or Neptune where the Kepler clock runs more slowly.
Second, this model does not explain what {\em halts} the accretion of
gas onto the planets.   
While it is possible the nebular gas was dispersed rapidly during the giant 
planet gas accretion phase, severe timing constraints (dispersal in $\lta 10^5$~years)
make this unlikely.  The most plausible explanation is that planetary gravitational tides
become strong enough to ``repel'' nearby gas once the planet is sufficiently massive
(\refcite{linpapIII}; see the next section).
Finally, it is not clear from the internal structure of the giant planets in our Solar System that
all four actually went critical.  The large gas masses of Jupiter and Saturn seem to require
their cores to have reached critical mass, but the smaller envelope masses of Uranus and Neptune
($M_{\rm env} \lta 0.5 M_{\rm core}$) imply these planets did not.
\tcite{shuevap} have discussed how photoevaporation of the disk
by the ultraviolet flux from the central star preferentially removes disk gas beyond
$\sim 10$~AU (the orbital distance of Saturn), which may explain the
smaller gas mass  reservoir available for the outer planets.
Recent calculations of the core-instability model that adopt more realistic planetesimal 
accretion rates (\refcite{pollack}) indicate the planets experience a long phase
($\sim 7\times 10^6$~year) after core runaway in which the envelope mass is smaller than the core mass
(see bottom panel in Fig.~\ref{coreinstability}).  If the nebula were removed during this long
phase of ``stalled'' gas accretion, the resulting planets would be similar to the outer giants.

\subsection{Tidal Interaction}
\label{sec:tides}

Gravitational interactions between orbiting mass points and disks has been considered by
many authors (\refcite{GT}; \citeauthor{linpapimpulse} \citeyear{linpapimpulse}; \refcite{wardhour};
\refcite{takeuchi}; see review by \refcite{linpapIII}).
The essential physics can be elucidated by using the impulse approximation (\refcite{linpapimpulse}).
As gas flows by the planet interior (exterior) to its orbital radius, the planet's
gravitational field causes a slight deflection in the gas streamlines causing the angular momentum of
the gas to decrease (increase).  The gravitational tidal interaction allows angular momentum
to be transferred from the gas on the inside to the planet and from the planet to the gas
on the outside.  Because orbital angular momentum increases outward in dynamically
stable disks (the Rayleigh criterion; \refcite{drazin}), 
gas interior (exterior) to the planet that
loses (gains) angular momentum in the tidal exchange
must move farther inward (outward).
In effect, the gravitational tides of the planet act to {\em repel} nearby gas.
In a fluid dynamical context (\refcite{GT}), the tides of the planet
excite spiral density waves that propagate away from the planet (in the absence of disk
self-gravity).  The waves are excited at Lindblad resonances, locations in the disk
where the natural frequency of radial oscillation (the epicyclic frequency) is an integer
multiple of the frequency at which the gas is forced by the 
planet.\footnote{For Kepler disks, the Lindblad resonances are at radii satisfying 
$\pm m (\Omega(r) - \Omega_{\rm p}) = \Omega(r)$, where $\Omega_{\rm p}$ is the
orbital frequency of the planet and $m$ is a positive integer.  The positive (negative)
sign is for inner (outer) resonances.}
Spiral density waves carry angular momentum, which is deposited in the gas as the waves are damped by
viscous stresses.  
Figure~\ref{disktorque} illustrates the situation schematically.
The total torque on the gas is found by summing the contributions from all the
Lindblad resonances; it is negative (positive) for the gas interior (exterior) to the planet.
The magnitude of the torque at either the inner or outer resonances, $T_{\rm tidal}$,
is approximately (\refcite{linpapIII})
\begin{equation}
\label{tidaltorque}
T_{\rm tidal} \approx f \, \Sigma \, \Omega_{\rm p}^2 \, r_{\rm p}^4 \, \left(\frac{r_p}{H}\right)^3 \,
\left(\frac{m_{\rm p}}{M_\ast}\right)^2,
\end{equation}
where $\Omega_{\rm p}$ is
the rotation frequency evaluated at the planetary radius $r_{\rm p}$, $H$ is the
vertical height of the disk and $f\approx 0.2$ is a numerical constant.
This torque is deposited in the vicinity of the planet unless the viscosity is
too small to damp the waves locally.
The {\em net} tidal torque,
which is the sum of the two nearly equal and opposite inner and and outer 
contributions, is difficult to calculate precisely but is roughly a factor
$H/r$ smaller than (\ref{tidaltorque}) (\refcite{ward86}).

\begin{figure}[htb]
\epsfxsize=\hsize
\epsfbox{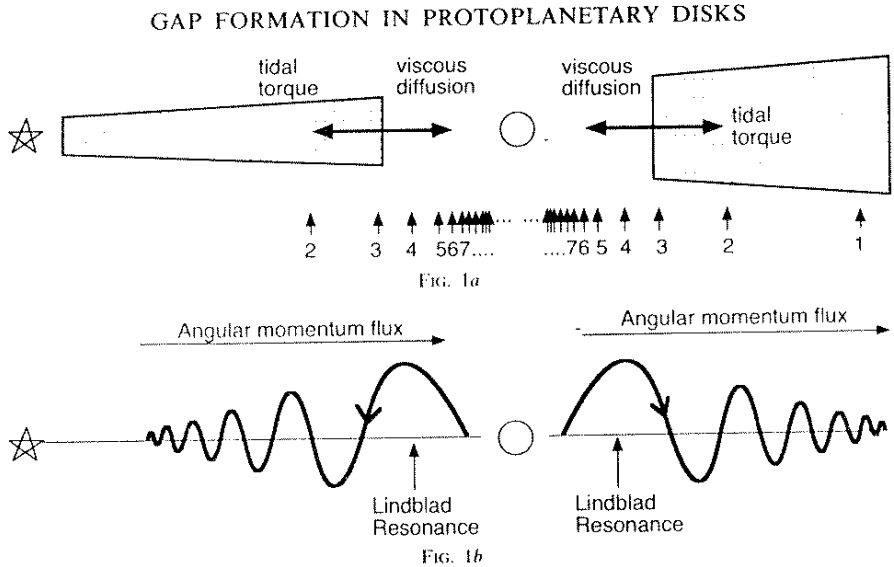}
\caption{Formation of gaps in disks from \tcite{takeuchi}.  
\textbf{Top Panel.} Schematic illustration 
shows that gaps are formed from a balance between tidal torques that repel gas from
the planet and viscous stresses that diffuse gas back into the gap.  The tidal torque is
actually deposited at the Lindblad resonances (indicated by arrows).  
\textbf{Bottom Panel.} Schematic illustration of wave propagation in disks.  Waves are excited
at the inner and outer 
Lindblad resonances and propagate away from the planet.  As they damp, the angular momentum
they carry is deposited into the gas as a tidal torque.  A net outward flux of angular momentum 
results.}
\label{disktorque}
\end{figure}

The action of the planetary tidal torques,
which tend to push gas away from the planet, is opposed by the viscous stress of
the gas, which tends to diffuse gas back toward the planet.
The viscous torque on the disk gas, $T_{\rm vis}$, is (\refcite{LBP})
\begin{equation}
\label{viscoustorque}
T_{\rm vis} = 3 \pi \Sigma \nu \, \Omega r^2 = 3 \pi \alpha \Sigma \Omega^2 r^4 \left(\frac{H}{r}\right)^2,
\end{equation}
using equations~(\ref{scaleht}) and (\ref{viscosity}) to get the rightmost form.
The criterion for the planet to open an annular gap
at its orbital radius is that
the gravitational tidal torque exceed the viscous torque, $T_{\rm tidal} \gta T_{\rm vis}$.
Using equations~(\ref{tidaltorque}) and (\ref{viscoustorque}) the minimum 
planet mass, $m_{\rm p}$, needed to open a gap is
(\refcite{takeuchi})
\begin{equation}
\label{gapminmass}
\frac{m_{\rm p}}{M_\ast} \; \gta \; \sqrt{\frac{3\pi}{f}} \, \left(\frac{H}{r}\right)^{5/2} \,
\alpha^{1/2}.
\end{equation}
For typical solar nebular parameters 
this gives a mass of $\sim 75\earth$, slightly less than the mass of Saturn.
It thus seems reasonable that Jupiter and Saturn (but not Uranus or Neptune unless the
disk was extremely cold) were able to form gaps
in the surface density distribution, reducing the local gas density to small values and
possibly terminating the rapid gas accretion phase.  
I note however that some recent hydrodynamical simulations of the gap formation process though have questioned
whether gap formation completely halts gas accretion onto the planet.
The simulations indicate that even if a low density
gap is formed, the planet remains connected
to the gap walls by arc-like streams through which gas can flow (\refcite{arty},
personal communication 1998).
Clearly far more work
is needed to determine whether the formation of a gap completely halts gas accretion, and
if not, what determines the final mass of the giant planets.

\subsection{Tidal Migration}
\label{sec:migrate}

An essential consequence of the tidal exchange of angular momentum between the planet
and disk is that a net torque can act on the planet, causing its orbital radius
to change with time.  This process is known as orbital or tidal {\em migration}
(\refcite{linpapmigrate}; \refcite{ward97}).
The angular momentum of the planet (on a circular orbit) 
is $m_{\rm p} \Omega_{\rm p} r_{\rm p}^2 = m_{\rm p}(G M_\ast r_{\rm p})^{1/2}$
and writing the {\em net} tidal torque exerted by the planet on the gas disk as $T_{\rm net}$, 
the orbital evolution of the planet is governed by 
\begin{equation}
\label{angmomevol}
\frac{d}{dt} \left(m_{\rm p} \Omega_p r_{\rm p}^2\right)  =  -\, T_{\rm net},
\end{equation}
or equivalently
\begin{equation}
\label{orbitevol}
\frac{d r_{\rm p}}{dt} = -\frac{2}{m_{\rm p} \Omega_{\rm p} r_{\rm p}} \; T_{\rm net}.
\end{equation}
The orbital migration timescale is
\begin{eqnarray}
\label{evoltime:nogap}
t_{\rm mig} &=&
 \frac{r_{\rm p}}{\vert dr_{\rm p}/dt\vert } = \frac{\Omega_{\rm p} r_{\rm p}^2}{2 T_{\rm net}},
\nonumber\\
& \sim & \frac{1}{\Omega_{\rm p}} \; \left(\frac{M_\ast}{m_{\rm p}}\right) \;
\left(\frac{M_\ast}{\Sigma r_{\rm p}^2}\right) \; \left(\frac{H}{r}\right)^2,\ \ \ \mbox{(no gap)}
\end{eqnarray}
where I have used equation~(\ref{tidaltorque}) reduced by $H/r$ to estimate the net torque
(\refcite{ward86}).  This equation is appropriate for lower mass planets that do not
tidally truncate the disk and form gaps.  For planets massive enough to form gaps, the 
surface density in their vicinity is reduced dramatically, and equation~(\ref{evoltime:nogap})
does not apply.  Numerical calculations of the combined gap formation and tidal migration
process (\refcite{linpapIII}) show that the inner and outer torques nearly cancel and
the evolution time becomes the {\em local} viscous diffusion time, equation~(\ref{viscoustime}),
applied at the orbit of the planet.
Figure~\ref{migrate} illustrates a self-consistent calculation
of gap formation and tidal migration in a protoplanetary disk.

\begin{figure}[htb]
\epsfxsize=\hsize
\epsfbox{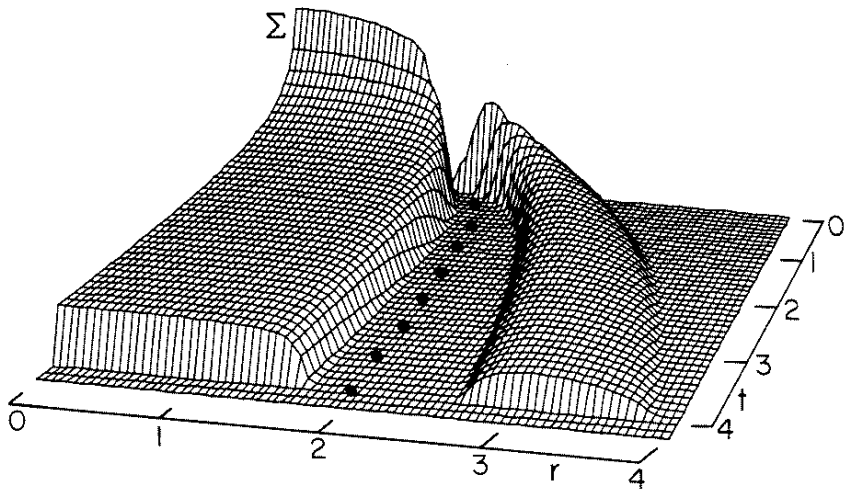}
\caption{Tidal migration of a protoplanet in an evolving nebula
from \tcite{linpapIII}.  Surface density, $\Sigma$, is plotted versus distance from
star, $r$, as a function of time, $t$.  All variables are scaled and dimensionless.
Notice the surface density decreases with time (due to accretion), and the outer disk radius
increases.  The position of the planet is marked by a dot. 
The mass of the planet is large enough that tides overcome viscous stresses,
and a  gap forms near the planet (the region where $\Sigma \longrightarrow 0$).
In this example, the planet is near the outer disk radius and migrates outward as 
it absorbs angular momentum from the inner disk.  Eventually the inner disk will be
accreted by the star, and the planet will migrate inward.  The migration timescale
is the viscous timescale of the disk.}
\label{migrate}
\end{figure}

The tidal migration timescale is 
surprisingly short.  Using values appropriate to
the \mmsn, the timescale (\ref{evoltime:nogap}),
appropriate for planets not forming gaps, 
is $t_{\rm mig} \approx (M_\ast/m_{\rm p})(r/{\rm AU})$~years, which is 
about $3\times10^{5}$~years for the Earth.   For planets massive enough to
form gaps, the ratio of the viscous
time at their orbital distance to the overall viscous time of the disk (having outer radius 
$R_{\rm D}$)
is $\sim (r_{\rm p}/R_{\rm D})^{3/2}$, from equation~(\ref{viscoustime}) assuming $H/r$ is
roughly constant.  Because Jupiter and Saturn were well within the outer radius of the primordial
solar nebula ($r_{\rm p} \ll R_{\rm D}$), 
their migration times were much shorter than the $\sim 10^7$ year lifetime of the disk as a whole.
Figure~\ref{sprmigrate} shows a calculation of the orbital migration of Jupiter and Saturn 
in which a migration time of $\sim  10^5$ years is found.

\begin{figure}[htb]
\epsfxsize=\hsize
\epsfbox{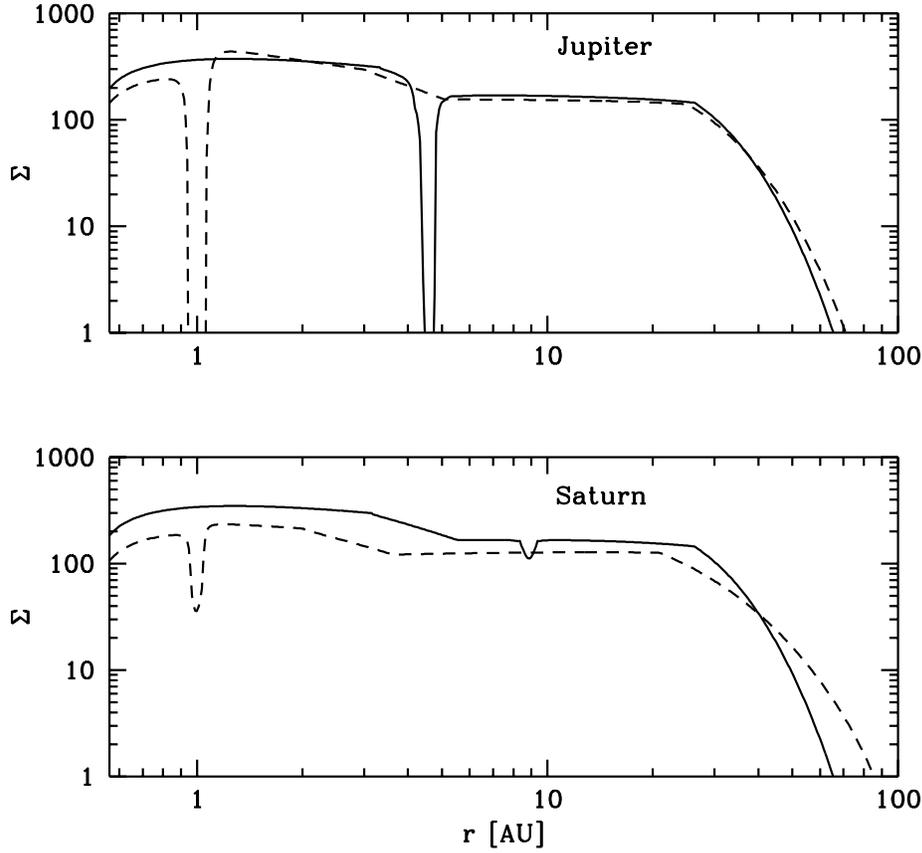}
\caption{Tidal migration in the solar nebula models of \tcite{rudlin}.
Surface density in g/cm$^2$ is plotted versus distance from the Sun in AU.
Jupiter is assumed to form instantaneously 
with its present mass and orbital radius at a time in the nebular evolution when
it is just massive enough to form a gap (eq.~[\ref{gapminmass}]). 
Saturn is assumed to form coevally.
\textbf{Top panel.} The solid curve is $10^3$ years after Jupiter formation showing
the gap (the region where $\Sigma$ drops to zero) is formed rapidly.  The dashed
curve is $4\times 10^{4}$~years later, showing Jupiter has migrated inward to 1~AU.
\textbf{Bottom panel.} The solid curve is also $10^3$~years after Saturn formation.
Saturn does not form a gap, but the surface density is reduced there by 30\%.
The dashed curve is $1.5\times 10^5$~years later.  
At this later time, the surface density in the gap is not zero but has been
reduced by a factor of 5.  The average inward migration speeds are about 40~cm/s
for either planet.}
\label{sprmigrate}
\end{figure}

\subsection{The Survival of Giant Planets}

Tidal migration offers a neat way to explain the recently
discovered Jupiter-mass planets having orbital distances of $\lta 0.1$~AU
from their stars (\refcite{marcybreview}; also see the review by Marcy in
this volume) if a mechanism can be found to {\em halt} the migration before
the planets are swallowed by the star.  The key is to find an outward torque 
that can counterbalance the inward tidal torque.
\tcite{linbodrich} have suggested that as
the planet nears the central star, it will raise tidal bulges in the
star that will transfer angular momentum from the spin of the star to the
planetary orbit.  This spin-orbit coupling can be effective at keeping the
planet beyond the corotation radius (the distance in the disk where the
Kepler orbital frequency equals the stellar rotation frequency).
These authors also suggest that a star with a sufficiently strong magnetic field
will have a magnetosphere that truncates the disk at a distance of several stellar radii.  Once
the planet migrates inside the inner disk edge, the inward tidal torques vanish because the
surface density is negligible.  Either mechanism can produce planets in orbits near the star.

\tcite{trilling} have considered a mechanism involving mass loss from the planet.
The size of the planetary Roche lobe ({\em c.f.} eq.~[\ref{hillsphere}]) is proportional to the 
distance from the star.  As a giant planet migrates inward, it can fill its Roche lobe,
leading to a transfer of mass from the planet to the star.  Conservation of angular momentum
of the system requires the planet to move outward.  Stable mass transfer is achieved
when the planet orbits at a distance where its radius equals its Roche radius.
Planets in small orbits near stars
can be produced if the disk is dissipated before the planet loses all its mass.

Our own Solar System with planets distributed from 0.4 - 30~AU
does not shown signs of appreciable orbital migration, however.
The question in our Solar System is to understand how the large orbital migrations 
apparently observed in other planetary systems were avoided. 
Because the migration timescale becomes the viscous timescale for planets massive
enough to form gaps, a giant planet can avoid appreciable migration if it
is born in a low viscosity nebula having evolutionary times
comparable to or longer than the disk dispersal time.
Before the planet has time to migrate far, the nebular gas is dispersed, and the migration
is halted.  Denoting the disk dispersal time by $t_{\rm disp}$, we can estimate 
from equation~(\ref{viscoustime}) that this requires the 
viscous alpha parameter to be $\lta (r/H)^2/(\Omega t_{\rm disp})$.  Using 
$r/H \approx 25$ and $t_{\rm disp}\sim 10^7$~years, I find $ \alpha \lta 10^{-4}$ is sufficiently
small that migration of Jupiter will not be severe.  It is not unreasonable that giant planets
form in a low viscosity environment.  The lower panel in Figure~\ref{coreinstability}
shows a model for the formation of Jupiter in which rapid gas accretion does 
not begin until $t \sim 8 \times 10^6$~years.  In this model, the newly formed Jupiter would find itself
in a nebular environment that had undergone significant evolution.  
The disk mass at that epoch could be quite small, with the corresponding low surface densities 
leading to small tidal torques and low migration rates.  In fact, given that the \mmsn\ 
contains only 10 -- 20 Jupiter masses, formation of the giant planets at a late epoch 
when the gas mass may be appreciably less than in the \mmsn\   may require the giant 
planets to accrete a large fraction of the remaining disk mass.

\subsection{Future Research Questions}
\begin{itemize}
\item Is the core-instability model the only way to make giant planets?
	\begin{itemize}
		\item Did the dispersal of the disk
		 prevent Uranus and Neptune from going critical?
	\end{itemize}
\item What determines the final masses and orbital distances of the giant planets?
	\begin{itemize}
		\item Does gap formation halt gas accretion onto giant planets?
		\item What prevented significant orbital migration from occurring in
		our Solar System?
	\end{itemize}
\item How does the planetary mass spectrum merge into the brown dwarf mass spectrum?
	\begin{itemize}
		\item Do brown dwarfs form in a fundamentally different manner
		than planets, {\em e.g.}, from collapse/fragmentation of the disk rather
		than from accumulation of dust?
	\end{itemize}
\end{itemize}

\par
\noindent
{\em Acknowledgments}

I would like to thank Charlie Lada and Nick Kylafis for inviting me to the
beautiful island of Crete for the second Star Formation summer school.  
This work has been supported in part by NSF AST--9157420 and NASA NAGW--5122.

\end{document}